%% file: main.tex
\documentclass[conference]{IEEEtran}
\usepackage{fancyhdr}
\usepackage{cite}
\usepackage{amsmath,amssymb,amsfonts}
\usepackage{algorithmic}
\usepackage{graphicx}
\usepackage{textcomp}
\usepackage[dvipsnames,table,xcdraw]{xcolor}
\usepackage{fontawesome5}
\usepackage{booktabs}
\usepackage{multirow}
\usepackage{subcaption}
\usepackage{url} 
\usepackage[ruled,vlined]{algorithm2e}
\usepackage{tikz}
\usepackage{amsmath,blkarray,bigstrut}

\newcommand*\numcircledtikz[1]{\tikz[baseline=(char.base)]{
            \node[shape=circle,draw,inner sep=1.2pt] (char) {#1};}}

\def\BibTeX{{\rm B\kern-.05em{\sc i\kern-.025em b}\kern-.08em
    T\kern-.1667em\lower.7ex\hbox{E}\kern-.125emX}}

\begin{document}

\title{PAL: A Variability-Aware Policy for Scheduling ML Workloads in GPU Clusters}
\author{\IEEEauthorblockN{Rutwik Jain, Brandon Tran, Keting Chen, Matthew D. Sinclair and
Shivaram Venkataraman}
\IEEEauthorblockA{Computer Sciences Department, University of Wisconsin-Madison\\
Madison, WI, United States of America\\
Email: \{rnjain, bqtran2, kchen346\}@wisc.edu,
\{sinclair, shivaram\}@cs.wisc.edu}}

\maketitle
\thispagestyle{fancy}
\lhead{}
\rhead{}
\chead{}
\lfoot{\footnotesize{SC24, November 17-22, 2024, Atlanta, Georgia, USA
\newline 979-8-3503-5291-7/24/\$31.00 \copyright 2024 IEEE}}
\rfoot{}
\cfoot{}
\renewcommand{\headrulewidth}{0pt}
\renewcommand{\footrulewidth}{0pt}

\input{abstract}

\begin{IEEEkeywords}
GPGPU; Cluster Scheduling; Machine Learning; Performance Variability; Power Management
\end{IEEEkeywords}

\maketitle %

\input{01_intro}
\input{02_background}
\input{03_design}
\input{04_methodology}

\input{05_evaluation}
\input{relWork}
\input{08_conclusion}

\input{acks}

\bibliographystyle{IEEEtran}
\bibliography{references}

\end{document}

%% file: abstract.tex
\begin{abstract}

Large-scale computing systems are increasingly using %
accelerators  %
such as GPUs to enable peta- and exa-scale levels of compute to meet the needs of Machine Learning (ML) and scientific computing applications.
Given the widespread and growing use of ML, including in some scientific applications, optimizing these clusters for ML workloads is particularly important.
However, recent work has demonstrated that accelerators in these clusters can suffer from %
\textbf{performance variability} and this variability can lead to resource under-utilization and load imbalance.
In this work we focus on how clusters schedulers, which are used to share accelerator-rich clusters across many concurrent ML jobs, can embrace performance variability to mitigate its effects.
Our key insight to address this challenge is %
to characterize which applications are more likely to suffer from performance variability and take that into account while placing jobs on the cluster.
We design a novel cluster scheduler, \textbf{PAL}, which uses performance variability measurements and application-specific profiles to improve job performance and resource utilization.
PAL also balances performance variability with locality to ensure jobs are spread across as few nodes as possible.
Overall, PAL significantly improves GPU-rich cluster scheduling: across traces for six ML workload applications 
spanning image, language, and vision models with a variety of variability profiles, PAL improves geomean job completion time by 42\%, cluster utilization by 28\%, and makespan by 47\% over existing state-of-the-art schedulers.

\end{abstract}

%% file: 01_intro.tex
\section{Introduction}
\label{sec:intro}

Artificial intelligence (AI) and machine learning (ML) have transformed society with significant improvements for a wide range of tasks~\cite{StateofAI22}.
This tremendous transformative effect has been enabled by a virtuous synergy of (1)~better hardware systems, (2)~larger datasets, and (3)~improved ML models (e.g., Transformers) and algorithms that further benefit from more efficient hardware and larger datasets.
ML is also increasingly impacting scientific applications~\cite{fan2021predicting, jumper2021highly, kates2019predicting}: ML models are either replacing or supplementing traditional computing methods in application domains like molecular dynamics (e.g., DeePMD~\cite{WangZhang2018-deepmd, ZengZhang2023-deepmd2}), protein folding (e.g., OpenFold2~\cite{openfold2}), and scientific AI models (e.g., AuroraGPT~\cite{Stevens2023-auroraGPT}).

However, meeting the computing needs of ML applications introduces new challenges.
With the slowing of Moore's Law and end of Dennard's Scaling,
large-scale systems are increasingly turning towards heterogeneous accelerators %
to scale performance, especially for ML workloads.
For example, large computing centers including cloud providers~\cite{FowersOvtcharov2018-brainwave,JouppiYoon2021-tpuv4,top500}
have deployed large accelerator-rich clusters that provide peta- or exa-scale levels of compute.
These systems often contain hundreds to tens of thousands of accelerators and are usually shared between many users.
Thus, \emph{cluster schedulers} need to handle large, accelerator-heavy ML workloads, %
while also aiming to reduce the time-to-solution of individual jobs and maintaining high resource utilization.

However, achieving high resource utilization for ML workloads is challenging in the face of performance variability of accelerators.
Prior studies~\cite{CoplinBurtscher2016-gpgpuPower, DeBardeleben-LBNL-EuroPar13, Fraternali-EEHPCVar-2018, JiaoLin2010-gpuPowerPerf, Scogland2015-pwrPerspectives, Tan-Gpuvar-2019, Sinha-SC22} have found that large clusters with accelerators like general-purpose GPUs (GPGPUs) exhibit significant \textit{performance variability}, both within %
and across machines (discussed further in Section~\ref{subsec:background-gpuVar}).
In our prior work we found that one of the main variability sources is power management (PM) in accelerators, which can lead to power and frequency variations across nodes~\cite{Sinha-SC22}. 
Performance variability also causes resource under-utilization for multi-GPU jobs since all of them must wait for the slowest one to complete due to the bulk synchronous programming (BSP) model used in data-parallel ML workloads~\cite{paszke2017-pytorch}.
Consequently, performance variability makes it challenging for ML workloads to achieve repeatable, high performance.

To overcome this challenge, our goal is to \textbf{harness and embrace performance variability}.
Specifically, we propose to redesign scheduling policies for GPU clusters to consider performance variability. 
Our key insight for designing a better policy comes from the fact that performance variability is \emph{application-specific}.
For example, prior work found that %
compute-intensive workloads such as training a ResNet-50 ML model had significant variability (22\% geomean variability, max 3.5$\times$).
Conversely, memory-intensive workloads such as PageRank had very low variability (1\%)~\cite{CoplinBurtscher2016-gpgpuPower, Sinha-SC22}.
Thus, we can create new policies that consider 
both the level of performance variability in a given cluster and how different applications are impacted by performance variability in that cluster.

We achieve our goal by (1) characterizing hardware performance variability by running a wide variety of single- and multi-GPU ML applications on the Texas Advanced Computing Center's (TACC)~\cite{tacc} Longhorn and Frontera clusters, (2) building application-specific performance variability profiles, and (3) designing new placement policies that utilize (1) and (2). 
Specifically, we propose a new job placement policy, \textbf{PM-First}, which considers PM-induced variability as the primary factor when assigning GPUs to jobs.
Our policy uses application profiles to give preferred GPUs to applications that are most sensitive to variability.
To ensure our policy can scale to handle large GPU clusters, we identify which GPUs exhibit similar performance variability and use K-Means clustering to group such GPUs together. 

While PM-First exclusively focuses on performance variability to make allocation decisions we find this sometimes results in sub-optimal schedules.
For example, if the accelerators with similar performance variability profiles are widely distributed across the cluster, scheduling to minimize or reduce performance variability may lead to significant overhead from inter-node communication (e.g., when weights are updated at the end of ML training epochs).
To address this challenge, we extend PM-First's algorithm to consider \textit{both} performance variability and locality when making its scheduling decisions.
We call this second approach \textbf{PAL} (\textbf{Performance Variability \& Locality)} since it aims to balance both concerns when scheduling multi-GPU jobs in the cluster.
PAL co-optimizes for both locality and variability by estimating the combined effect of both for every possible GPU allocation for a given job. 
However, selecting the best allocation considering both factors can be expensive for a large cluster.
Thus, we propose an efficient mechanism where we construct and traverse a locality-variability matrix (L$\times$V matrix).
The L$\times$V-matrix is succinct and its size is bound by the number of locality levels in the cluster (i.e., the network topology hierarchy) and the number of K-Means clusters used for grouping PM scores.
We show that this allows PAL to behave the same as PM-First for variability-sensitive jobs, and make locality-first allocations for jobs that need to prioritize network communication.

Given the importance of ML workloads running in large-scale clusters, prior work has also examined scheduling policies for ML  workloads~\cite{gao2022-survey,Gu-Tiresias-NSDI19,Qiao-Pollux-OSDI21,Narayanan-pop-sosp21}.
However, almost all of these scheduler designs are agnostic to GPU variability and assume that iso-architecture GPUs deliver equal performance.
One notable exception is Gavel~\cite{Narayanan-Gavel-osdi20}, which considers performance heterogeneity but only across different accelerator architectures in heterogeneous clusters.
Thus, to the best of our knowledge, our work is the first to make cluster schedulers aware of iso-architecture GPU performance variability.
We discuss this and other related work further in Section~\ref{sec:relWork}.

To evaluate the efficacy of PM-First and PAL, we integrated them into Blox~\cite{agarwal2023blox}, a state-of-the-art toolkit that supports many modern scheduling and placement schemes for ML workloads. %
While our policies can potentially also benefit other types of workloads such as scientific computing, their evaluation is left for future work.
We compare our PM-First and PAL placement policies with widely used locality-based placement policies from Tiresias~\cite{Gu-Tiresias-NSDI19} and Gandiva~\cite{Xiao-Gandiva-OSDI18}, which perform job packing. We also evaluate our placement with different scheduling policies, including FIFO, LAS, and SRTF. 
Overall, PM-First and PAL significantly improve cluster scheduling for a variety of ML workload traces from prior work~\cite{Subramanya-Sia-Sosp23,Mohan-Synergy-OSDI22}.
For example, PM-First improves geomean 99th percentile job completion time (JCT) by 40\%, average JCT by 40\%, utilization by 26\%, and makespan by 44\% over Tiresias. 
By balancing both performance variability and locality, PAL further improves on PM-First.
Compared to Tiresias, PAL improves geomean 99th percentile JCT by 41\%, average JCT by 42\%, and makespan by 47\%.
PM-First and PAL benefits are especially large when workloads have a large proportion of multi-GPU jobs -- the increasingly common case as ML model sizes continue growing.
Thus PAL's benefits are likely to scale further with future workloads.
Finally, we validate our approach using a 64-GPU TACC Frontera cluster and find that PAL outperforms Tiresias by 24\% in terms of average JCT. %

%% file: 02_background.tex
\section{Background}
\label{sec:background}

\subsection{GPU Performance Variability}
\label{subsec:background-gpuVar}

GPUs in large clusters such as datacenters and supercomputers exhibit \emph{variability} in performance, despite having the same underlying architecture and being similarly configured. 
Prior studies~\cite{Scogland2015-pwrPerspectives,CoplinBurtscher2016-gpgpuPower,DeBardeleben-LBNL-EuroPar13,Fraternali-EEHPCVar-2018,Tan-Gpuvar-2019,Sinha-SC22} have examined and characterized performance, power, and temperature variability among GPUs at scale. 
For example, our prior work profiled 5 large, GPU-rich clusters with 400$-$27000 identical GPUs and identified significant performance %
variability, both within %
and across machines.
Similar to prior work~\cite{CoplinBurtscher2016-gpgpuPower}, we found that different applications exhibit different extents of performance variability at scale.
In particular, compute-bound workloads %
like ResNet-50 see much more variation (22\% geomean, up to 3.5$\times$) than memory-intensive workloads like PageRank (1\% geomean).

Performance variability occurs for a number of reasons in these systems.
Static effects at the hardware level such as process variation and die binning cause inherent manufacturing variability among GPUs, while power and temperature limits and associated PM algorithms cause dynamic variation. 
In HPC systems, non-uniformity in cooling across nodes can also cause thermal throttling, increasing performance variability~\cite{Ostrouchov-Titan-2020}.
This variability is not transient either: performance variability is consistent over time, and ill-performing GPUs are consistently ill-performing~\cite{Sinha-SC22}.
Moreover, not all the performance variability can be explained by temperature differences or cooling sources, and the variability is consistent across different days of the week, times of day, and GPU vendors.
Moreover, as discussed in Section~\ref{sec:intro} massively parallel workloads that synchronously utilize multiple GPUs are bottlenecked by the worst-performing GPU -- reducing cluster underutilization and load imbalance, hurting overall throughput and efficiency of these large-scale systems.
Consequently, this application-specific performance variability affects application performance across runs on the same cluster~\cite{DeBardeleben-LBNL-EuroPar13, Sinha-SC22} and is a growing problem for accelerator-rich systems.

\begin{figure}[tb!]
    \centering
    \includegraphics[width=0.55\linewidth]{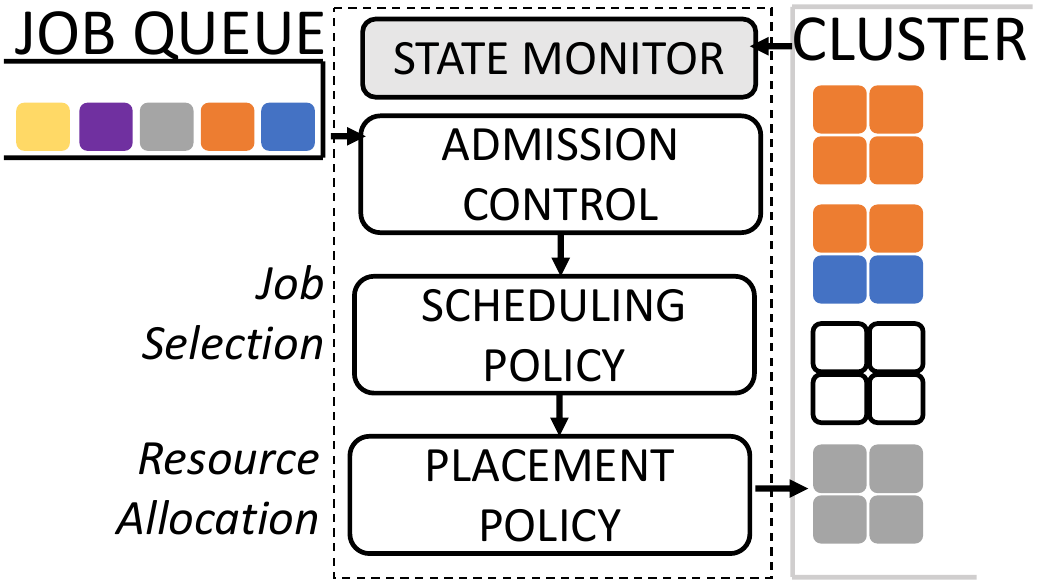}
    \caption{Modular view of Blox job scheduling.}
    \label{fig:sched-overview}
\end{figure}

\subsection{Cluster Scheduling}
\label{subsec:backgnd-sched}

ML workloads are increasingly being run on shared, GPU-rich clusters.
Thus, efficient scheduling is necessary to minimize ML training time and efficiently utilize cluster resources.
Collectively, the cluster scheduler must decide which job(s) to schedule at a given time and what resources they should be given, respectively.
The architecture of most cluster schedulers can be broken up into modules that determine job admission, job selection, and resource allocation~\cite{agarwal2023blox}.


Figure~\ref{fig:sched-overview} provides a broad overview of these modules.
All incoming jobs are put into a queue and admitted based on an admission control policy. 
Schedulers typically admit jobs that do not adversely impact the performance of currently running jobs and do not violate resource constraints~\cite{Gu-Tiresias-NSDI19, Subramanya-Sia-Sosp23}. 
The scheduling policy receives these accepted, active jobs to schedule in each epoch or \emph{scheduling round}.
The active jobs are then assigned priorities and reordered appropriately by the \textit{scheduling policy}, as per the scheduling objective.
Finally, this ordered job queue is forwarded to the \textit{placement policy}, which determines what resources (e.g., GPU(s)) should be allocated to a job based on the state of the cluster that the scheduler actively monitors.
While the scheduling policy selects which jobs to run at a given epoch, the placement policy determines which GPUs to run them on. 
Job selection is largely orthogonal to our work, since we are focused on incorporating GPU performance variability into allocation decisions.
Thus we focus on the scheduler's placement policy. 
As discussed further in Section~\ref{sec:relWork}, current state-of-the-art placement policies are agnostic to GPU performance variability. 
When allocating GPU resources to jobs, they assume that iso-architecture GPUs deliver equal performance.
Therefore, GPU placement policy designs that harness GPU performance variability information 
when making allocation decisions are needed.

%% file: 03_design.tex
\section{Design}
\label{sec:design}

We propose placement algorithms that make GPU variability a first-class citizen when determining GPU job allocations. 
We leverage a key insight from previous work~\cite{CoplinBurtscher2016-gpgpuPower, Sinha-SC22}: \textit{GPU variability is application-specific; e.g., compute-bound workloads exhibit higher variability than memory-bound ones.}
This implies that memory-bound jobs can use GPUs that widely vary for compute-intensive jobs without significant performance loss, and allow compute-intensive jobs to utilize GPUs with similar variability. 

\begin{figure}[tb!]
    \centering
    \includegraphics[width=0.75\linewidth]{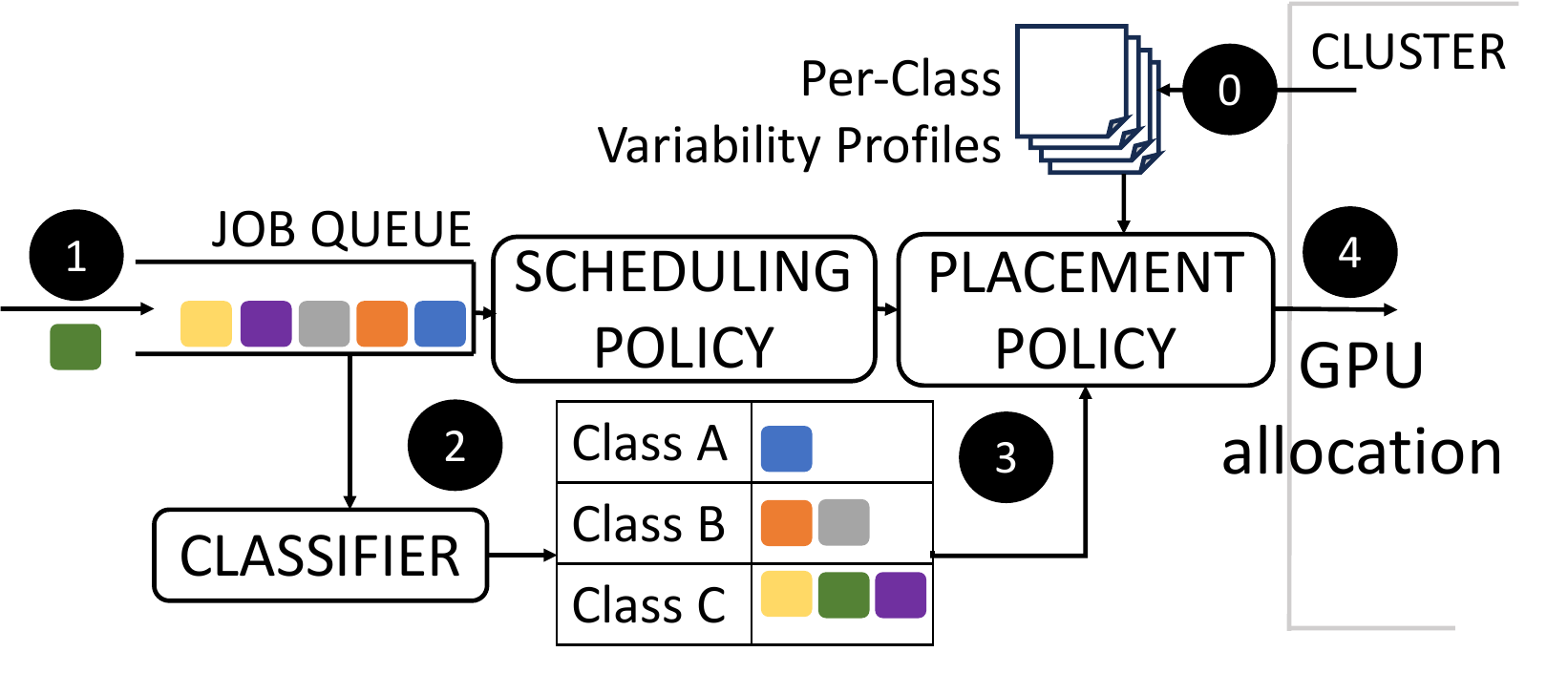}
    \caption{Variability-aware scheduling overview.}
    \label{fig:pal-overview}
\end{figure}

Figure~\ref{fig:pal-overview} provides an overview of our approach. 
We perform offline profiling to gather variability profiles from the target cluster and pass this information to our placement policy \numcircledtikz{0}.
Since performance variability is application-specific, we create and use an application classifier (Section~\ref{subsec:design-classifier}) that groups similarly behaving applications into a small number of classes, based on compute intensity 
\numcircledtikz{2}.
Figure~\ref{fig:pal-overview} shows various jobs grouped into three classes, A, B, and C, with A representing the most compute-intensive, and C representing the most memory-bound applications. 
Section~\ref{subsec:design-classifier} further discusses the classifier.
Arriving jobs are put into the job queue \numcircledtikz{1} and the classifier tags these applications with a suitable class \numcircledtikz{2}. 
The scheduler receives active jobs and assigns them priorities based on its scheduling objective. %
This sorted queue of jobs is forwarded to the placement policy, which determines what resources should be allocated to a job. 
Our placement policy receives the job's performance class \numcircledtikz{3}, allowing it to make application-specific decisions; and profiled variability data \numcircledtikz{0} which it uses to make variability-aware allocations \numcircledtikz{4}.
We propose two algorithms for placement: PM-First (Section~\ref{subsec:pmfirst-design}) and PAL (Section~\ref{subsec:pal-design}).

\subsection{Classification Layer}
\label{subsec:design-classifier}

GPU clusters concurrently run multiple jobs, and new ones arrive frequently.
Moreover, new applications, particularly ML models, are emerging frequently and often change the footprint of workloads run on the cluster~\cite{comp_bw_scaling, ShoeybiPatwary2019-megatronlm, NaffzigerBeck2021-ryzen, JouppiYoon2021-tpuv4}.
It is infeasible to profile such a large range of applications, especially at scale, to measure performance variability across thousands of GPUs. 
To reduce the number of required application profiles, we use a classifier that groups similarly behaving applications into a small number of classes.
This significantly reduces the amount of profiling data we need to collect.

\begin{figure}[tb!]
    \centering
    \includegraphics[width=\linewidth]{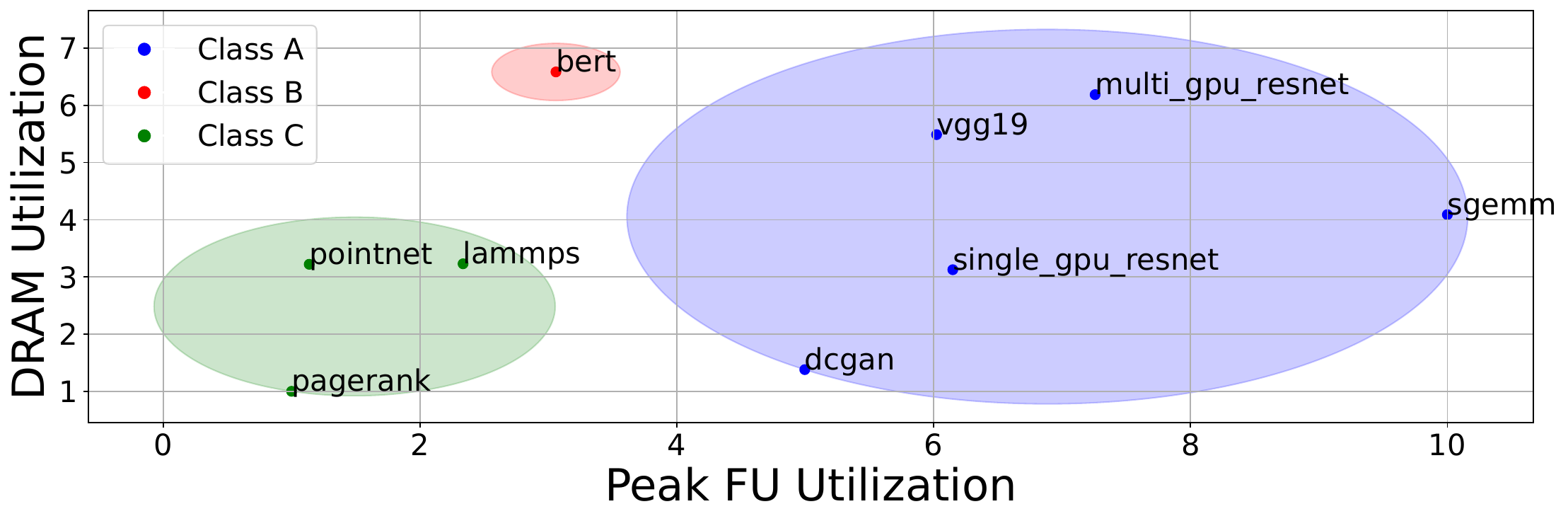}
    \caption{Classification of applications using 2-dimensional clustering over the $\text{Util}_{\text{DRAM}} \times \max(\text{Util}_{\text{FU}}) $ space.}
    \label{fig:classification}
\end{figure}

To classify the variability of the applications we study (Section~\ref{subsec:methodology-appsConfig}), we leverage prior work's application classification scheme~\cite{Adolf-Fathom-IISWC16, Guerreiro-DVFSaware}.
Like Guerreiro, et al.~\cite{Guerreiro-DVFSaware}, we use \texttt{nsight compute}~\cite{NsightCompute} to measure workloads’ DRAM utilization ($\text{DRAM}_{\text{Util}} $) and Peak Functional Unit utilization ($\text{PeakFU}_{\text{Util}}$).
\begin{equation*}
    \text{DRAM}_{\text{Util}} = \frac{\text{DRAM}_{\text{Bandwidth}}}{\text{DRAM}_{\text{PeakBandwidth}}}\times 10
\end{equation*}
\begin{equation*}
    \text{PeakFU}_{\text{Util}} = 
    \max_{i \in \text{FuncUnits}}
    \large\{ \text{FU}_{\text{Util}}^i \large\}
\end{equation*}
Here, $\text{FuncUnits}$ refers to different compute components of the GPU, namely, single precision, double precision, texture, special and tensor function units. For an application with $T$ unique kernel types, we compute individual function unit utilization as: 
\begin{equation*}
     \text{FU}_{\text{Util}}^i = \frac{\sum_T \text{kernel\_runtime} \times \text{kernel\_util}^i }{\sum_{T} \text{kernel\_runtime}}\times 10
\end{equation*}
Note that \texttt{nsight compute} reports utilization metrics in a [0,10] range.

Each workload corresponds to a point in the 2-dimensional $\text{DRAM}_{\text{Util}} \times \text{PeakFU}_{\text{Util}}$ space.
Figure~\ref{fig:classification} shows the applications we consider in this 2D space.
Then we perform $K$-Means clustering to obtain ordered classes. 
$K$ can be appropriately set to choose the number of classes for the classifier.
For example, by setting $K=3$ in Figure~\ref{fig:pal-overview} jobs are assigned one of three classes: A, B, or C, where A is the most sensitive to variability and C is the least sensitive to variability.
For a new application or an application with different input parameters/datasets, we profile the application and assign it to the cluster it is closest to in the 2D space. 
We retain this three-class example to illustrate the design of our policies in subsequent sections.

\subsection{PM-First Placement Policy}
\label{subsec:pmfirst-design}

PM-First gives PM-induced variability first-order precedence when assigning GPUs to jobs.
To carry out PM-First allocations, we associate a PM-Score with each GPU, which indicates how slow or fast the GPU is relative to the median GPU in the cluster.
The PM-Scores for a GPU are computed for each job class since each class has a different variability profile. 
We perform application-specific variability profiling by running an application on each GPU to collect performance metrics and use these profiled performance values normalized to the median GPU on the cluster. 
For example, a PM-Score of 1.5 for a GPU $g$ means that a job's iteration time will be slowed down by 50\% running on $g$ compared to the median performing GPU.
Since the PM-Score values for a GPU correspond to normalized execution time on the GPU, we refer to GPUs with lower PM-Scores as well-performing GPUs.

\begin{figure}[bt!]
    \centering
    \includegraphics[width=0.9\linewidth]{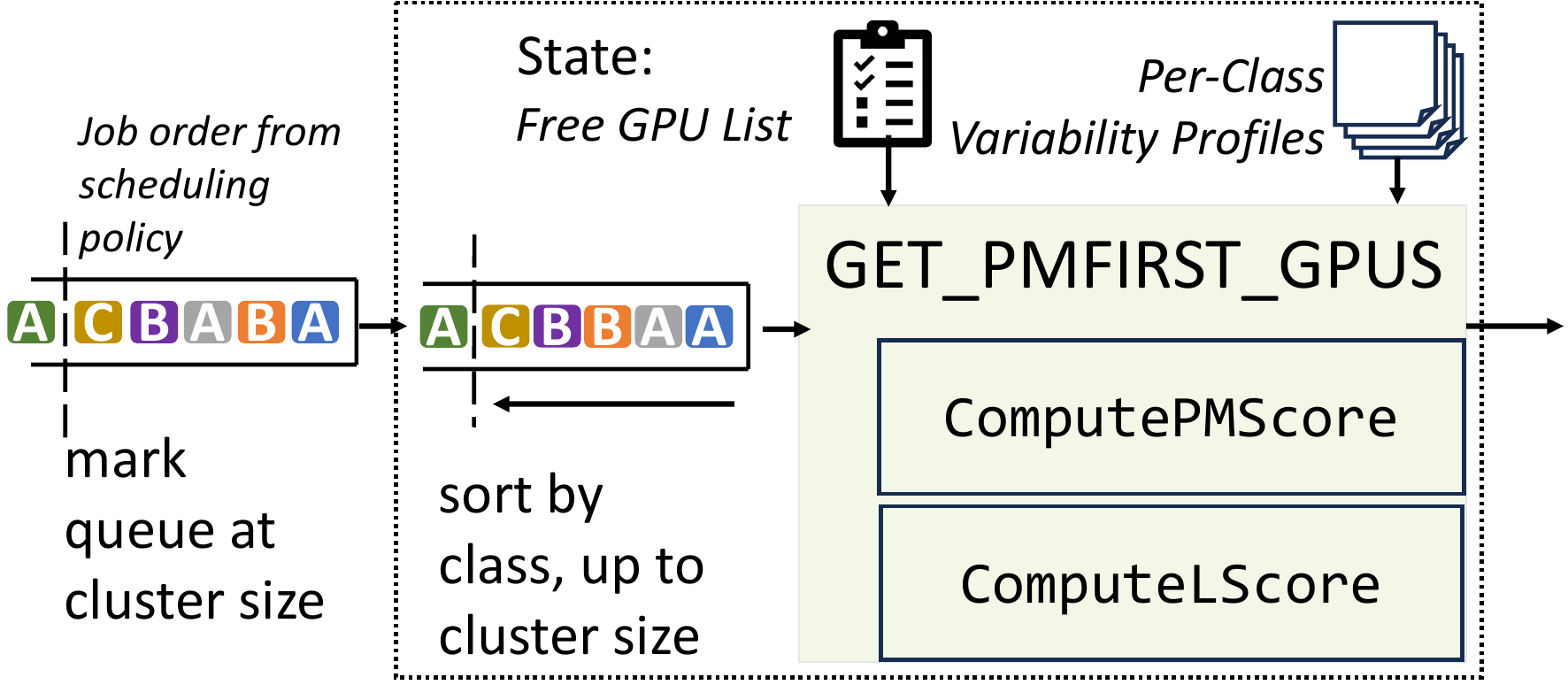}
    \caption{PM-First Placement Policy with job queue reordering based on placement priority.}
    \label{fig:pmfirst-policy}
\end{figure}

Since class A applications are most sensitive to variability, we assign class A jobs the highest placement priority, then class B, and so on. 
PM-First placement follows a greedy algorithm where the GPUs with the lowest PM-Scores 
are assigned to jobs with the highest placement priority.
This placement priority is different from the scheduling priority produced by the scheduling policy.
Figure~\ref{fig:pmfirst-policy} shows an example of a 6-job queue sorted by the scheduling policy and sent to our placement policy for GPU assignment.
The class A jobs in the queue have higher placement priority since we want to provide a larger set of well-performing GPUs to these jobs. 
However, we must respect the scheduling priorities set by the scheduling policy.
To ensure this we mark the job queue when the sum of GPU demands of jobs in the queue exceeds the cluster size.
In Figure~\ref{fig:pmfirst-policy}, the GPU demand exceeds cluster size after the first 5 jobs.
Thus, the scheduling policy must guarantee that these 5 jobs be scheduled in this round.
Therefore, we only sort this truncated queue by job class, allowing compute-intensive class A jobs to select GPUs first, then class B, and so on.
This prevents an incoming class A job (marked green in Figure~\ref{fig:pmfirst-policy}) from getting dispatched out of turn.
By separating scheduling priority and placement priority, we honor the scheduling policy's guarantee of which jobs to service in a given round, while still allowing GPU allocations to be done using a class-based priority.

Algorithm~\ref{alg:pmfirst} shows how PM-First allocates resources for a given job $j$ with a GPU demand $N_j$. 
We compute the PM-Scores $V_i$ for each GPU $i$ using the variability profile for job $j$'s class (\texttt{ComputePMScore}). 
Then we sort the free list of available GPUs by their PM-Scores, from best to worst.
The policy picks the first $N_j$ GPUs to satisfy the GPU demand for job $j$. 
This process continues for the next job(s) in the modified job schedule. 

\input{listings/pmfirst}

Using fine-grained variability information when scheduling a large-scale system with thousands of GPUs could be expensive. 
For instance, Oak Ridge National Lab's (ORNL's) Summit supercomputer has over 27000 GPUs, and assigning a PM-Score to each of these GPUs and tracking them during run time adds significant overhead to the scheduler.
Thus, we use K-Means clustering for each class to bin GPU variability values into a set of PM-Scores. 

\begin{figure}[tb!]
    \centering
    \includegraphics[width=0.7\linewidth]{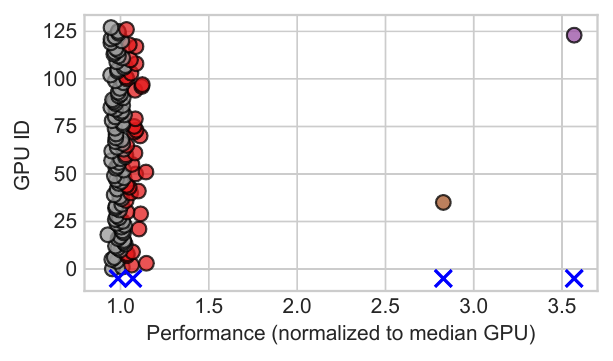}
    \caption{Example showing clustering on a 128-GPU cluster for a class A application. A blue cross marks each bin's centroid.}
    \label{fig:clustering}
\end{figure}

Figure~\ref{fig:clustering} shows our clustering method employed on the variability profile of a compute-intensive application, ResNet-50, for a 128-GPU cluster.
The x-axis shows average iteration time normalized to the median GPU of the cluster.
Most GPUs belong to the first 2 clusters close to the median, while some outliers are more than $2.5\times$ slower than the median.
We associate the PM-score for all GPUs in a given bin with the PM-score of the centroid.
For example, in Figure~\ref{fig:clustering} all GPUs of the gray cluster get assigned a PM-Score of 0.99.

If we pick very small values for $K$ (the number of bins produced by clustering), we lose fine-grained variability information, and PM-First cannot differentiate between GPUs that deliver different performances. 
Conversely, very high $K$ values overestimate the impact of variability, making PM-First more selective in picking GPUs than necessary. 
Thus, we need to determine optimal $K$ values for each class. 
We select the optimal $K$-value using the standard silhouette score method~\cite{Rousseeuw-SilScore-1987}. 
However, since the variability data has some extreme outliers, particularly for compute-intensive applications, this adversely impacts silhouette coefficients.
We separate $>3\sigma$ outliers when computing silhouette scores and sweeping through $K$ from 2 to 11. 
We select the $K$ value that gives silhouette scores as close to +1 as possible for all bins so that we get distinct and relatively well-separated bins.
We similarly determine an optimal $K$ value for the set of outliers.
Note the placement policy does not ignore $>3\sigma$ outliers; they are only removed for the silhouette score analysis.
These extreme outliers are assigned their own PM-score equal to the GPU's normalized performance.

\subsection{PAL Placement Policy}
\label{subsec:pal-design}

While the PM-First policy factors variability into its decision-making, it ignores communication overheads that may occur due to ineffective packing. 
Thus, PM-First works well for applications that are sensitive to variability (e.g., class A), but not for those that are less impacted by variability 
(typically class C). 
Accordingly, our PAL placement policy co-optimizes for locality and variability by observing their combined effects, ensuring that PAL prioritizes either packing or variability depending on what is more important.

\subsubsection{Combined Slowdown and $L\times V$ Matrix}
\label{subsubsec:pal-design-slowdown}

Large-scale systems typically have a flat network topology without much over-subscription.
For example, TACC Frontera uses a Mellanox interconnect in a fat tree topology with an 22/18 oversubscription~\cite{tacc}. 
For such systems, where there is no complex routing or multiple hops between nodes of the same layer, we use a simplified locality model.  
A multi-GPU job incurs a performance penalty $L_{across}$ if its allocation spills across nodes and suffers no performance degradation if the allocation is within a node ($L_{within} = 1.0$). 
If a job is running with a set of GPUs $G$ and the GPUs  are spread across more than one node, then the job's modified iteration time is: %
\begin{equation}
    \label{eqn:iter_time_pal}
    t_{iter} = L_{across} \times \max_{g \in G}(V_g) \times t^{orig}_{iter} 
\end{equation}
where $t^{orig}_{iter}$ is the job's original iteration time, 
as specified in Section~\ref{sec:background}, %
$V_g$ is the variability or PM-Score of the $g$th GPU and $L_{across}$ is the inter-node locality penalty.
Section~\ref{sec:methodology} provides more details on estimating a cluster's locality penalty.

To get better performance for jobs, we need to minimize the combined slowdown due to both variability and locality penalties. 
We denote this as the LV-Product:
\begin{equation*}
    \min \text{LV-Product} = \min ( L_1 \times \max_{g \in G}V_g )
\end{equation*}

To minimize this product, we construct an $L \times V$ matrix for each job class at design time based on the profiled variability and the locality penalty.
For example, consider the following $L \times V$ matrix with 4 bins for PM-Scores $V_1 = 0.89, V_2 = 0.94, V_3 = 1.06$, and $ V_4 = 2.55$, and a constant inter-node locality penalty $L_{across} = 1.5$. 

{{\footnotesize
\begin{equation*}
\mathbf{L\times V}= 
\begin{blockarray}{ccccc}
V_1 (0.89) & V_2 (0.94) & V_3 (1.06) & V_4 (2.55) & \\
\begin{block}{[cccc]c}
  0.89 & 0.94 & 1.06 & 2.55 & L_{\text{within}}(1) \\
  1.34 & 1.41 & 1.59 & 3.88 & L_{\text{across}}(1.5) \\
\end{block}
\end{blockarray}
\end{equation*}
}}

The matrix entries represent the LV-Product we want to minimize.  
Each entry corresponds to a possible allocation scenario. 
We ``traverse" this matrix from smallest LV-Product to largest, making job allocations to minimize the LV-product (Algorithm~\ref{alg:pal}, line 3).
In this example, the $L \times V$ matrix traversal order would be:
$ (1,0.89) \rightarrow (1,0.94) \rightarrow (1,1.06) \rightarrow (1.5,1.34) \rightarrow (1.5,1.41) \rightarrow (1.5,1.59) \rightarrow (1.5,3.88)$.
In other words, unlike PM-First, PAL allows non-packed allocations only if the first three variability bins cannot provide a packed allocation to service this job. However, PAL prefers a distributed allocation over allocating GPUs from bin 4 which has a very high PM-Score ($V_4 = 2.55$).
Moreover, the $L \times V$ matrix is class specific traversal orders are often unique per-class.
This allows PAL to make PM-First decisions for variability-sensitive jobs, and make locality-first allocations for jobs that need to prioritize packing.

Algorithm~\ref{alg:pal} details the steps to perform this $L \times V$ traversal.
For a job requesting $N_j$-GPUs on a cluster size $N$, there are $^NC_{N_j}$ possible GPU allocations. 
Our inter-node cost model for locality allows us to reduce this search space, since jobs with GPU demand $N_j >$ \texttt{NUM\_GPUS\_PER\_NODE} must request multiple nodes and pay the inter-node locality penalty of splitting across nodes. 
PAL schedules all such jobs using the PM-First policy (Algorithm~\ref{alg:pal}, lines 23-25).
A job $j$ requesting $N_j$ GPUs, where $1 < N_j \leq \texttt{NUM\_GPUS\_PER\_NODE}$, traverses the allocation in two ways: 

\begin{enumerate}
    \item $(L_{within}, V_i)$ allocations: PAL needs to prioritize packing while making sure that PM-Score for the allocation is $\leq V_i$. We filter out the free list of GPUs with $V \leq V_i$ and then try to enumerate strictly packed (within-node) allocations within this free list by enumerating all possible packed $N_j$-sets of GPUs and finding the set with the least variability.
    \item $(L_{across}, V_i)$ allocations: we filter out the free list of GPUs with PM-Score $\leq V_i$ and sort them by PM-Score. Since locality cost is acceptable to incur in this state, we pick the first $N_j$ GPUs from this sorted list. 
\end{enumerate}

\input{listings/pal}

This $L \times V$ matrix traversal only occurs for jobs that require \texttt{NUM\_GPUS\_PER\_NODE} or fewer number of GPUs. 

%% file: listings/pmfirst.tex
{
\begin{algorithm}[tb!]
\caption{\texttt{GET\_PMFIRST\_GPUS} \\ PMFirst Selection Algorithm}
\label{alg:pmfirst}
\footnotesize
\SetAlgoLined
\KwIn{Free GPU List $G_{free}$ 
\\ Job Class $C_j$
\\ Job GPU Demand $N_j$ 
\\ Variability Profile $V_{profile}$ }
\KwOut{GPU Allocation $Alloc$}

\SetKwFunction{GetPMFirstGPUs}{GET\_PMFIRST\_GPUS}
\SetKwFunction{ComputePMscore}{ComputePMscore}
\SetKwFunction{ComputeLocalityScore}{ComputeLocalityScore}
\SetKwFunction{Sort}{Sort}
\SetKwFunction{Alloc}{MarkGPUsInUse}

\SetKwProg{Fn}{Function}{:}{}

\nl\Fn{\GetPMFirstGPUs{$G_{free}$, $C_j$,$N_j$,$V_{profile}$ }}{
    \nl\tcp{Get per-GPU PM-Scores $V_i$ corresponding to job class}
    \nl\ForEach{gpu $i \in G_{free}$}{
        \nl$V_i$ $\leftarrow$
        \ComputePMscore{$V_{profile}$, $C_j$}\;
    \nl}
    \nl\tcp{Sort free GPU list by PM-Score, from best to worst}
    \nl$G_{free}$ $\leftarrow$ \Sort($G_{free}$, $V_i$, descending)\;
    \nl\tcp{Select the top $N_j$ number of GPUs}
    \nl$Alloc$ $\leftarrow$ $G_{free}$[:$N_j$]\;
    \nl\tcp{Remove from free GPU list}
    \nl\Alloc();
    
    \nl\Return{allocation}\;
}
\end{algorithm}
}

%% file: listings/pal.tex
{{
\begin{algorithm}[tb!]
\caption{\texttt{PAL\_PLACEMENT} \\ PAL Selection Algorithm}
\label{alg:pal}
\footnotesize
\SetAlgoLined
\KwIn{Free GPU List $G_{free}$, 
\\ Job Class $C_j$
\\ Job GPU Demand $N_j$ 
\\ $L\times V$ matrix
}
\KwOut{GPU Allocation $Alloc$}

\SetKwFunction{PALPlacement}{PAL\_PLACEMENT}
\SetKwFunction{Traverse}{traverse}
\SetKwFunction{Filt}{filt}
\SetKwFunction{FindValidNodes}{FindValidNodes}
\SetKwFunction{GenerateCombinations}{generateCombinations}
\SetKwFunction{GetMinV}{GetMinV}
\SetKwFunction{Allocate}{allocate}
\SetKwFunction{GetPMFirstGPUs}{getPMFirstGPUs}

\SetKwProg{Fn}{Function}{:}{}

\nl \Fn{\PALPlacement{$G_{free}$, $C_j$, $N_j$, $L\times V$}}{
    \nl \If{$1 < N_j \leq \texttt{NUM\_GPUS\_PER\_NODE}$}{
       \nl  \For{($L_i, V_i$) \textbf{in} \Traverse{$L\times V$}}{
           \nl  \If{$L_i = L_{within}$}{
                \nl \tcp{Filter GPUs with PM Scores better or equal to $V_i$}
                \nl $G_{filt}$ $\leftarrow$ $G_{free}$[$V \leq V_i$]\;
                \nl \tcp{Enumerate potential packed allocations}
                \nl $n$ $\leftarrow$ \FindValidNodes($G_{filt}$, $N_j$)\;
                \nl $PackAlloc$ $\leftarrow$ GenerateCombos($^{n}C_{K}$)\;
                \nl \tcp{Return one with lowest PM Score}
                \nl $Alloc$ $\leftarrow$ \GetMinV{$PackAlloc$}\;
                \nl \Return{$Alloc$}\;
            \nl }
            \nl \ElseIf{$L_i = L_{across}$}{
                \nl \tcp{PM-First allocation}
                \nl $G_{filt}$ $\leftarrow$ \Filt{free\_gpu\_list, $V \leq V_i$}\;
                \nl $Alloc$ $\leftarrow$ $G_{filt}$[:$N_j$]\;
                \nl \Return{$Alloc$}\;
            \nl}
        \nl}
    \nl}
    \nl\Else{
        \nl\tcp{PM-First allocation}
        \nl$Alloc$ $\leftarrow$ \GetPMFirstGPUs{}\;
        \nl\Return{$Alloc$}\;
    }
}

\end{algorithm}
}}

%% file: 04_methodology.tex
\section{Methodology}
\label{sec:methodology}

\subsection{System}
\label{subsec:methodology-sys}

We run experiments on both a physical cluster and in simulation.
All experiments use Blox~\cite{agarwal2023blox}, an open-source modular toolkit that uses Python and gRPC~\cite{grpc} to support scheduler implementation and testing. 
The physical cluster experiments are performed on TACC's Frontera supercomputer~\cite{tacc}.
Frontera is a mineral-oil cooled GPU subsystem with 360 NVIDIA Quadro RTX 5000 GPUs.
Each node has 4 GPUs, with 16GB memory per GPU.
We run physical cluster experiments on an 16 node (64 GPU) testbed. 
We also use larger trace-based simulations to evaluate the behavior of our policies with varying cluster sizes, traces, and scheduling policies. 
Table~\ref{tab:exptlist} summarizes the key experiments and system details.

\input{tables/exptlist}

\subsubsection{Baseline Placement Policies}
\label{subsubsec:methodology-sys-place}
We evaluate PM-First and PAL's performance relative to two baselines:
(1) Packed or soft-consolidated placement tries to minimize the number of nodes a job is packed on to reduce communication; and 
(2) Random or Scattered placement samples a random subset from the free list of GPUs in order to prevent thermal hotspots through unbalanced GPU usage, increase device lifespan, and prioritize performance of CPU-to-GPU communication. However, random placement can sacrifice performance for workloads sensitive to GPU-to-GPU communication.

We further consider two flavors of each of these policies -- Sticky and Non-Sticky.
In Sticky placement, active jobs cannot be migrated to a different allocation and must continue to run with the same (``sticky") set of GPUs they first get assigned, until the jobs either complete or get preempted through priority lowering. 
Thus, the Sticky placement policy only re-allocates GPUs to a job once the job moves from suspended to active state. 
Sticky allocations minimize checkpointing overheads that occur due to migration, but these are typically negligible relative to the overall job run-time.
We chose these as our baselines because most state-of-the-art schedulers use one of them. Specifically, we compare against the following configurations: 
\begin{enumerate}
    \item \textbf{Tiresias}~\cite{Gu-Tiresias-NSDI19}: performs Packed-Sticky placement
    \item \textbf{Gandiva}~\cite{Xiao-Gandiva-OSDI18}: performs Packed-Non-Sticky placement
    \item \textbf{Random-Sticky}
    \item \textbf{Random-Non-Sticky} 
\end{enumerate}
Other schedulers also use some variants of these policies.
For example, Themis~\cite{Mahajan-Themis-OSDI20} also uses Packed placement, while Amaral, et al. and HotGauge use Random placement~\cite{Amaral-Topology-SC17, Hankin-HotGauge-IISWC21}.
In the remainder of this paper we use \textbf{Tiresias} to mean Packed-Sticky placement and  \textbf{Gandiva} to mean Packed-Non-Sticky placement.

Our \textbf{PAL} and \textbf{PM-First} placement policies are both Non-Sticky to ensure jobs %
can 
migrate to better GPUs in each scheduling round. 

\subsubsection{Scheduling Policies}
\label{subsubsec:methodology-sys-sched}

We evaluate three schedulers PM-First and PAL placement can be attached to:

\noindent    
\textbf{First-In-First-Out} (\textbf{FIFO}) scheduler: a well-known greedy approach that prioritizes jobs in order of arrival. 

\noindent
\textbf{Tiresias/Least Attained Service} (\textbf{LAS}): implements LAS scheduling with two-level priority queuing~\cite{Gu-Tiresias-NSDI19}.

\noindent
\textbf{Shortest Remaining Time First} (\textbf{SRTF}): performs preemptive shortest job first scheduling.

While we present simulation results with all three scheduling policies, we use Tiresias for the TACC Frontera cluster runs.
Section~\ref{subsec:synergy-sim-eval} compares how the behavior of our placement policy changes when the scheduler is varied. 

\subsection{Workloads and Cluster Configuration}
\label{subsec:methodology-appsConfig}

\subsubsection{Simulations}
\label{subsubsec:methodology-appsConfig-sim}

To compare our placement policies against the baselines (Section~\ref{subsubsec:methodology-sys-place}) in simulation we use two sets of workload traces:

\noindent
\textbf{Sia-Philly workloads}~\cite{Subramanya-Sia-Sosp23} sample jobs from Microsoft's publicly available Philly cluster production traces~\cite{Jeon-Philly-2019}.
Sia derives eight traces of 160 jobs each, submitted over an 8 hour window at a job arrival rate of 20 jobs/hr. 
We use these traces in the same cluster configuration Sia used: a simulated 16 node system with 4 GPUs per node (64 GPUs). 
40\% of Sia trace jobs are single-GPU jobs, and the largest multi-GPU jobs request up to 48 GPUs.
We report average metrics such as Job Completion Time (JCT) across these 160 jobs, consistent with prior work that used these traces~\cite{Subramanya-Sia-Sosp23, Qiao-Pollux-OSDI21}.

\noindent
\textbf{Synergy workloads} preserve the Philly trace's~\cite{Jeon-Philly-2019} GPU demand and use a Poisson distribution of arrival times to vary job arrival rate.
Synergy traces have a higher proportion of single-GPU jobs ($> 80\%$) than Sia-Philly traces. 
To evaluate Synergy's steady state benefits, like prior work we simulate Synergy on a larger 64 node, 4-GPU per node cluster (256 GPUs) and report average metrics for job IDs 2000 to 3000.

\subsubsection{Cluster Evaluation on TACC Frontera}
\label{subsubsec:methodology-appsConfig-frontera}

Table~\ref{tab:modellist} lists the jobs we run for our real cluster experiments on the TACC Frontera cluster.
As with the Sia simulations, we track average JCT metrics across all 160 jobs. 
In an extended version~\cite{Jain24PAL-arxiv} we present results with additional HPC workloads.

\input{tables/blox-job-models}

\subsection{Estimating PM penalty}
\label{subsec:pmpenalty-estimation}

We perform variability profiling to estimate per-GPU, per-class PM penalties. 
We use two sets of variability profiles from different systems, one from TACC's Longhorn cluster (NVIDIA V100 GPUs), used for simulations, and one from TACC's Frontera cluster (NVIDIA Quadro RTX 5000 GPUs) used in both simulation and cluster experiments. 
These profiles are generated by running the same benchmark on all GPUs of the cluster and collecting metrics for kernel duration/iteration runtime. 
As discussed in Section~\ref{subsec:design-classifier}, we profile one representative application for each class. 
Specifically, we profile ResNet50 for Class A, BERT for Class B and PageRank for Class C. 
Table~\ref{tab:summary-benchmarks} summarizes the applications we ran for profiling and their input configurations on each cluster.
We used NVIDIA's \texttt{nsight compute}~\cite{NsightCompute} to collect performance metrics in milliseconds (ms) for these applications. 
Since \texttt{nsight compute} can only provide 1 sample every ms, we configured our input sizes to ensure that kernel durations were larger than 1ms on the respective GPUs.

\begin{figure}[tb!]
    \centering
    \includegraphics[width=0.7\linewidth]{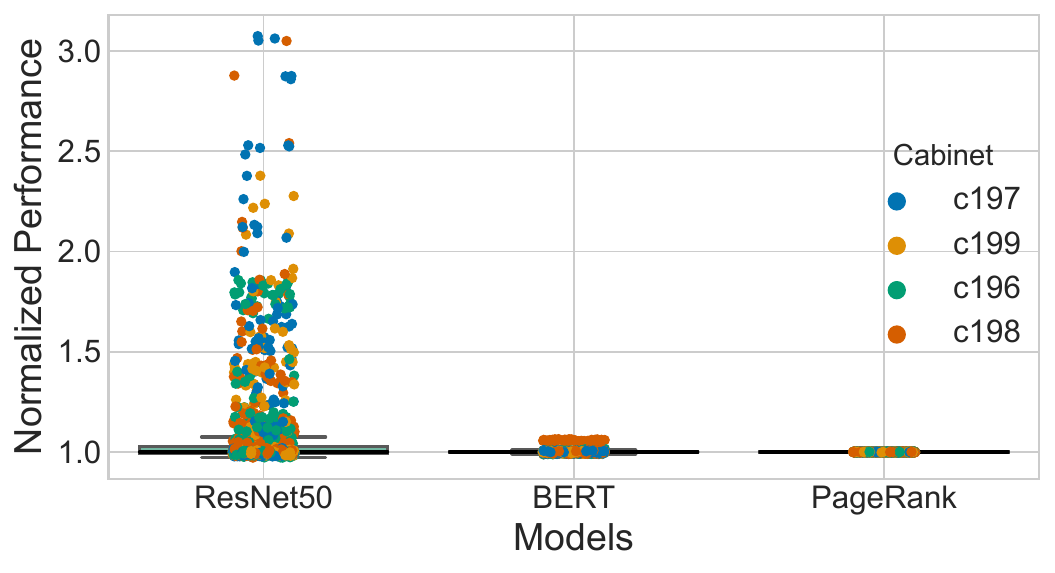}
    \vspace{-1ex}
    \caption{Normalized Frontera cluster performance variability.}
    \label{fig:perf_front}
\end{figure}

\begin{figure}[tb!]
    \centering
    \includegraphics[width=0.7\linewidth]{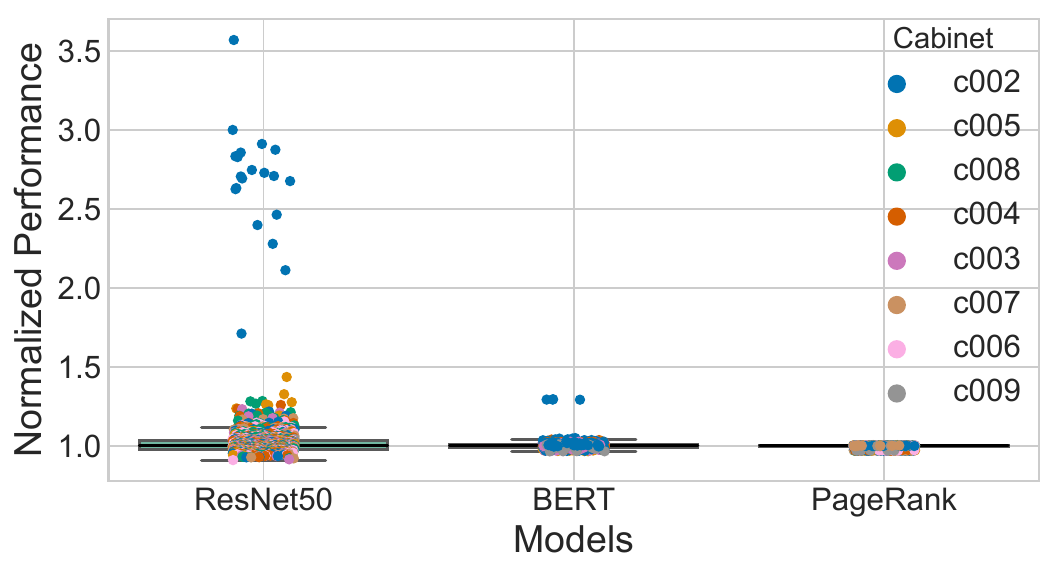}
    \vspace{-1ex}
    \caption{Normalized Longhorn cluster performance variability.}
    \label{fig:perf_long}
\end{figure}

Figure~\ref{fig:perf_front} and Figure~\ref{fig:perf_long} show each application's GPU performance normalized to the application's median  observed duration on the cluster.
When simulating an $N$-GPU cluster, we discretely, randomly sample this profiling data without repetition to obtain $N$ PM penalty values for each class and assign them to each GPUs.

\input{tables/variability-benchmarks}

For the Frontera testbed implementation, we index into the variability profile using GPU UUID obtained from \texttt{nvidia-smi}~\cite{nvidia-smi} to get the exact PM penalty value from the variability profiling data.
Profiling for a large number of GPUs could be time-consuming, so our variability profiles are static -- they are generated at design time and remain constant throughout. 

\subsection{Locality Penalty}
\label{subsec:methodology-localPenal}

We estimated the inter-node locality penalty $L_{across}$ by profiling the iteration time for a 4-GPU ResNet-50~\cite{job-model-resnet50} job with a batch size of 64 run on all 4 GPUs of a single node versus an 8 GPU ResNet-50 job with a batch size of 128 running on two nodes. 
The ratio of the average iteration time aggregated from the two profiles gives us an estimate for locality penalty. 
Using this method we initially estimated a locality penalty to be $1.7$ on TACC Frontera.
We use this locality penalty in our Synergy simulations, but study different locality penalty values to evaluate how our policies fare in other systems with different costs of distributing jobs across nodes.
Our physical experiments showed that inter-node communication costs are not as high on Frontera, and are also model-dependent.
Hence, we estimate per-model locality penalties from our physical cluster experiments, and use these in simulation experiments in Sections~\ref{subsec:real-cluster-eval} and~\ref{subsec:sia-philly-eval}.

%% file: tables/exptlist.tex
{\footnotesize
\begin{table}[tb!]
\centering
\caption{List of experiments used in evaluation.}
\label{tab:exptlist}
\begin{tabular}{@{}llll@{}}
\toprule
\textbf{\begin{tabular}[c]{@{}l@{}}Workload\\ Trace\end{tabular}} & \textbf{\begin{tabular}[c]{@{}l@{}}Cluster Size\\ (NumGPUs)\end{tabular}} & \textbf{Experiment}  & \textbf{\begin{tabular}[c]{@{}l@{}}Eval. \\ Section\end{tabular}} \\ \midrule
\textbf{Sia-Cluster}~\cite{Subramanya-Sia-Sosp23}  
& 64  
& Testbed Evaluation 
& \ref{subsec:real-cluster-eval} \\ \midrule

\textbf{Sia-Philly}~\cite{Subramanya-Sia-Sosp23,Jeon-Philly-2019} 
& 64   
& \begin{tabular}[c]{@{}l@{}}Baseline Simulation\\ Varying Locality Penalty\end{tabular}  
& \ref{subsec:sia-philly-eval} \\ \midrule

\textbf{Synergy}~\cite{Mohan-Synergy-OSDI22}     
&  256 
& \begin{tabular}[c]{@{}l@{}}Varying Job Load\\ Varying Schedulers\end{tabular} 
& \ref{subsec:synergy-sim-eval} \\ 
\bottomrule
\end{tabular}
\vspace{-2ex}
\end{table}
}

%% file: tables/blox-job-models.tex
{{\footnotesize
\begin{table}[tb!]
\centering
\caption{Models used in real cluster evaluation.}
\label{tab:modellist}
\begin{tabular}{@{}lllll@{}}
\toprule
\textbf{Task}     & \textbf{Model}     & \textbf{Dataset}      & \textbf{{\begin{tabular}[c]{@{}l@{}}Batch\\ Size\end{tabular}}} & \textbf{Class}\\ \midrule
Image    & PointNet~\cite{job-model-pointnet}  & ShapeNet~\cite{dataset-shapenet}     & 32      & Class C      \\
Image    & vgg19~\cite{job-model-vgg19}         & ImageNet2012~\cite{dataset-imagenet} & 32    & Class A         \\
Vision   & DCGAN~\cite{job-model-dcgan}        & LSUN~\cite{dataset-lsun}         & 128        & Class A    \\
Language & BERT~\cite{job-model-bert,megatron-lm}          & WikiText~\cite{dataset-wikitext}    & 64         & Class B    \\
Image    & ResNet-50~\cite{job-model-resnet50} & ImageNet2012~\cite{dataset-imagenet} & 32 & Class A \\ 
Language    & GPT2~\cite{megatron-lm}        & Wikitext~\cite{dataset-wikitext} & 128& Class B \\ \bottomrule
\end{tabular}
\end{table}
\vspace{-1ex}
}}

%% file: tables/variability-benchmarks.tex
{\footnotesize
    \begin{table}[tb!]
  \caption{Applications profiled for PM penalty estimation}
  \centering
  \begin{tabular}{@{}llll@{}}
    \hline
    \multirow{2}{*}{\textbf{Benchmark}} & \multirow{2}{*}{\textbf{Input Size}} & \textbf{Clusters}  \\
    & & \textbf{Observed} \\ \hline
    \multirow{2}{*}{\textbf{ResNet50}\cite{ResNet-pyTorchRef}} & Train. Set: 1.2M images & Longhorn  \\
    & Batch size: 64  & Frontera \\ \hline
    \multirow{2}{*}{\textbf{BERT}\cite{DevlinChang18-bert}} & Train. Set: 30K words & Longhorn  \\
    & Batch size: 64  & Frontera \\ \hline
    \multirow{2}{*}{\textbf{PageRank}\cite{CheBeckmann2013-pannotia}} & $643994\times643994$ & Longhorn \\ 
    & $659033\times659033$  & Frontera \\ \hline    
  \end{tabular}
  \label{tab:summary-benchmarks}
  \end{table}
}

%% file: 05_evaluation.tex
\section{Evaluation}
\label{sec:eval}

\subsection{Real Cluster Experiments}
\label{subsec:real-cluster-eval}

\begin{figure}[tb!]
    \centering
    \includegraphics[width=0.8\linewidth]{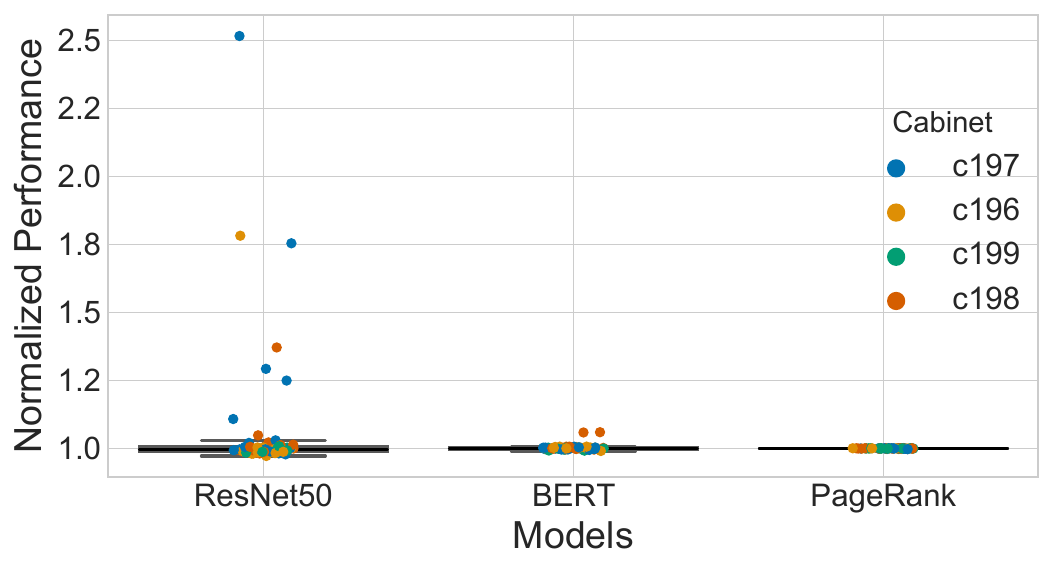}
    \vspace{-2ex}
    \caption{TACC Frontera 64-GPU testbed variability profiles.}
    \label{fig:cluster-variability}
\end{figure}

To ensure our reported gains from Blox (Sections~\ref{subsec:sia-philly-eval}-\ref{subsec:synergy-sim-eval}) are representative and realistic, we first compare \textbf{PAL} to \textbf{Tiresias}, the best performing baseline, on a physical 64-GPU cluster.
Figure~\ref{fig:cluster-variability} shows the variability profile of this 64-GPU testbed.
Since we are running these experiments on a physical cluster, we use the exact PM penalties for the 64 GPUs that we used in our experiments.

\begin{figure}[tb!]
    \centering
    \includegraphics[width=0.6\linewidth]{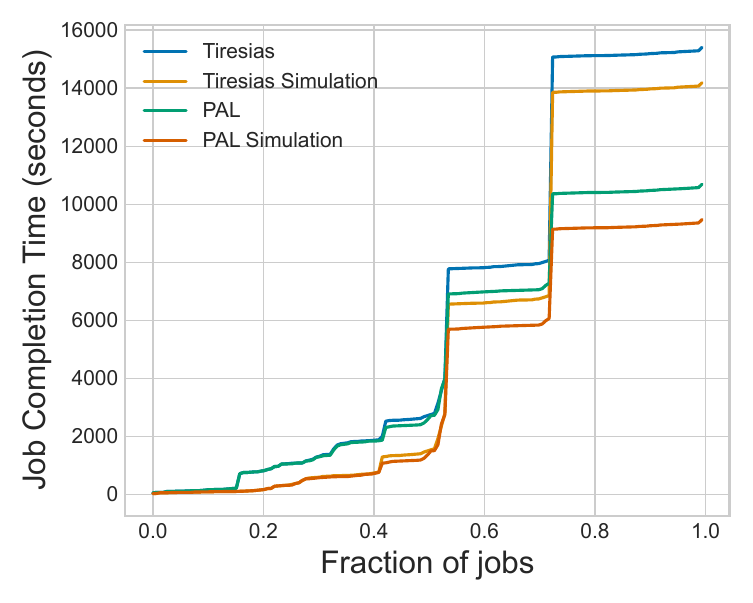}
    \vspace{-1ex}
    \caption{Cluster and simulation cumulative Job Completion Time distributions.}
    \label{fig:cluster-cdf}
\end{figure}

Figure~\ref{fig:cluster-cdf} shows the cumulative distribution of JCTs for the physical cluster experiments as well as the corresponding simulation experiments (Sections~\ref{subsec:sia-philly-eval}-\ref{subsec:synergy-sim-eval}) for the same trace.
Broadly, the cluster and simulation CDFs align fairly well for both policies.
Overall, PAL reduces average JCT and makespan by 24\%  and 27\%, respectively, over Tiresias' placement policy, whereas simulation predicts a 28\% benefit given the testbed's variability and locality penalties.
These results demonstrate the benefit of our approach: even versus the best performing baseline, PAL's ability to effectively exploit both locality and variability improves performance.

\input{tables/cluster-vs-sim-results}

Table~\ref{tab:cluster-res} summarizes \textbf{PAL} and \textbf{Tiresias}'s physical cluster average JCT results.
Although Figure~\ref{fig:cluster-cdf} showed that the cluster and simulator JCT distributions exhibit the same trends, the difference between the average JCT metric in physical and simulated clusters is 11-14\% for both policies.
Upon further investigation of job iteration times, we found that the profiled PM-scores of GPUs in node 0 (c196-071) for class A jobs were much lower ($\sim$8$\times$) than the PM penalties actually experienced by these jobs on the cluster.
This accounts for the majority of the gap between cluster and simulation JCTs, with a few large jobs running slower on the cluster than in simulation.
Consequently, subsequent jobs spend additional time waiting for resources, increasing their JCTs as well. 
This highlights the need for periodic re-profiling of the cluster, or dynamic online updates to GPU PM-Scores to more accurately reflect the cluster's variability characteristics.
Although it does not contribute to the difference between simulated and physical cluster JCTs, the profiled variability of the 64 GPU subset of Frontera (Figure~\ref{fig:cluster-variability}) is lower than Frontera's overall profile (Figure~\ref{fig:perf_front}) for Class A jobs.
Specifically, the 64 GPUs we ran on have 6\% variability for ResNet-50 (Class A), versus 13.3\% in the overall profile.
Nevertheless, our results demonstrate that PAL still significantly improves average JCT over Tiresias.

\begin{figure}[tb!]
    \centering
    \includegraphics[width=\linewidth]{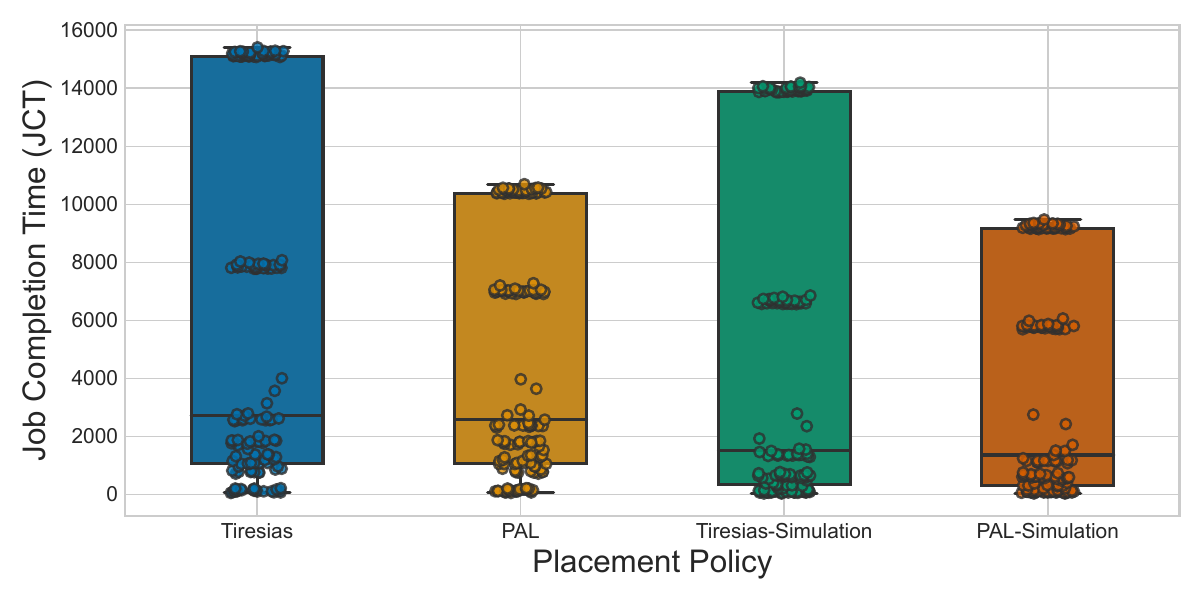}
    \caption{Physical cluster and simulated \textbf{Tiresias} and \textbf{PAL} JCT boxplots.}
    \label{fig:cluster-jct-boxplots}
    \vspace{-2ex}
\end{figure}

Overall, these results demonstrate both the viability of our technique on real GPU clusters and the fidelity of our simulator, which we next utilize to evaluate a wider range of configurations in simulation.

\subsection{Sia-Philly Simulations}
\label{subsec:sia-philly-eval}

\begin{figure}[tb!]
    \centering
    \includegraphics[width=\linewidth]{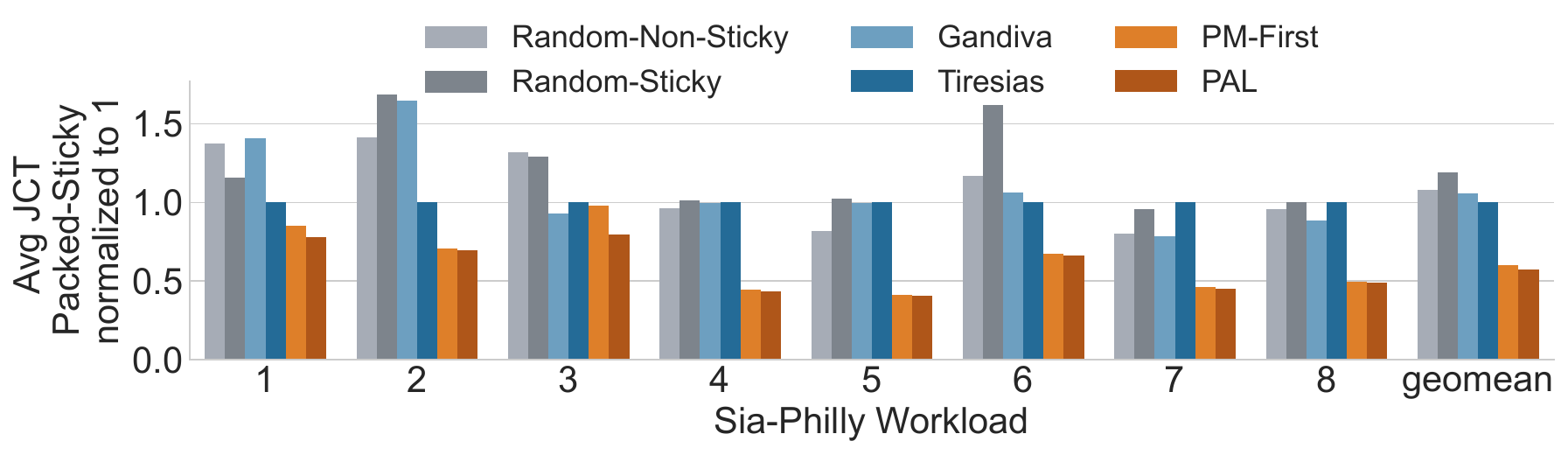}
    \caption{Average JCT, normalized to \textbf{Tiresias} placement, for different Sia-Philly workloads on a 64-GPU cluster with FIFO scheduling policy.}
    \label{fig:sia-baseline}
\end{figure}

Next we evaluate Sia-Philly traces in a simulated 64-GPU cluster with FIFO scheduling using PM penalty profiles from TACC's Longhorn cluster. 
Further configuration details were described in Section~\ref{subsubsec:methodology-appsConfig-sim},~\ref{subsec:methodology-localPenal}, and~\ref{subsec:pmpenalty-estimation}. 
Figure~\ref{fig:sia-baseline} compares the average Job Completion Time (JCT), normalized to \textbf{Tiresias} or Packed-Sticky placement.

\noindent
\textbf{Random \& Packed Policies}: Since random selection increases the likelihood that the allocated GPUs are ill-performing in terms of variability and/or packing, the \textbf{Random} baseline policies often suffer in terms of performance (e.g., workloads 1, 2, 3, 6, and geomean JCT results).
Conversely, the packed baselines, namely \textbf{Tiresias} and \textbf{Gandiva}, improve performance by minimizing the number of nodes a job is packed on, thereby reducing the inter-node locality cost for multi-GPU jobs.

\noindent \textbf{Sticky \& Non-Sticky Variants}: For most workloads (1, 2, 4, 5, 6, geomean), \textbf{Tiresias}' sticky placement outperforms  \textbf{Gandiva}'s non-sticky GPU selection.
This happens because non-sticky placement picks GPUs every scheduling round, increasing the likelihood of jobs picking GPUs with worse PM-Scores for some rounds during their runtime.
With Sticky placement jobs that pick GPUs with high PM-Scores get consistent slowdowns throughout their runtime.
However, because Sticky placement keeps the same GPUs throughout execution, there are fewer such jobs compared to Non-Sticky placement.

\begin{figure}[tb!]
    \centering
    \begin{subfigure}[b]{0.2\textwidth}  
    \includegraphics[width=\linewidth]{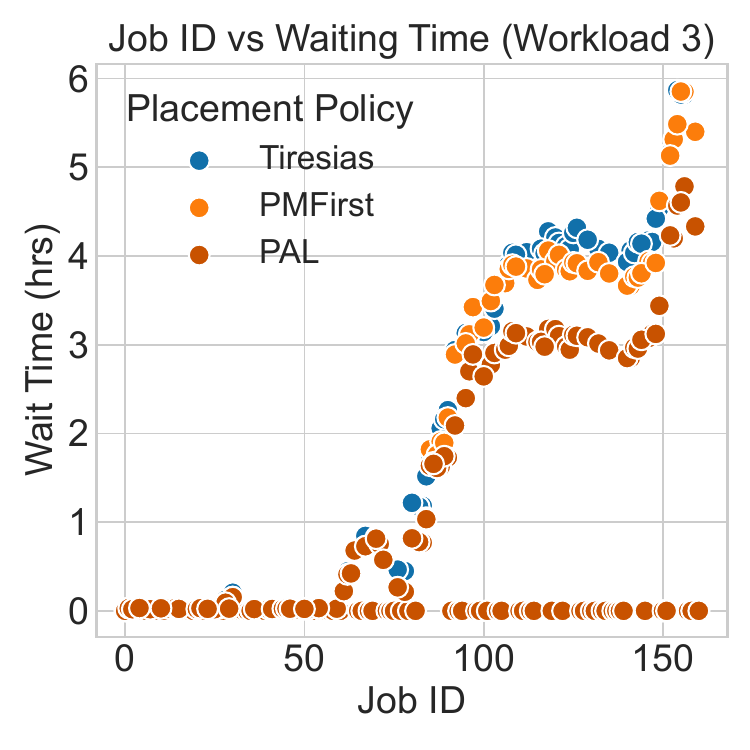}
    \label{subfig:w3}
    \end{subfigure}
    \begin{subfigure}[b]{0.2\textwidth}       \includegraphics[width=\linewidth]{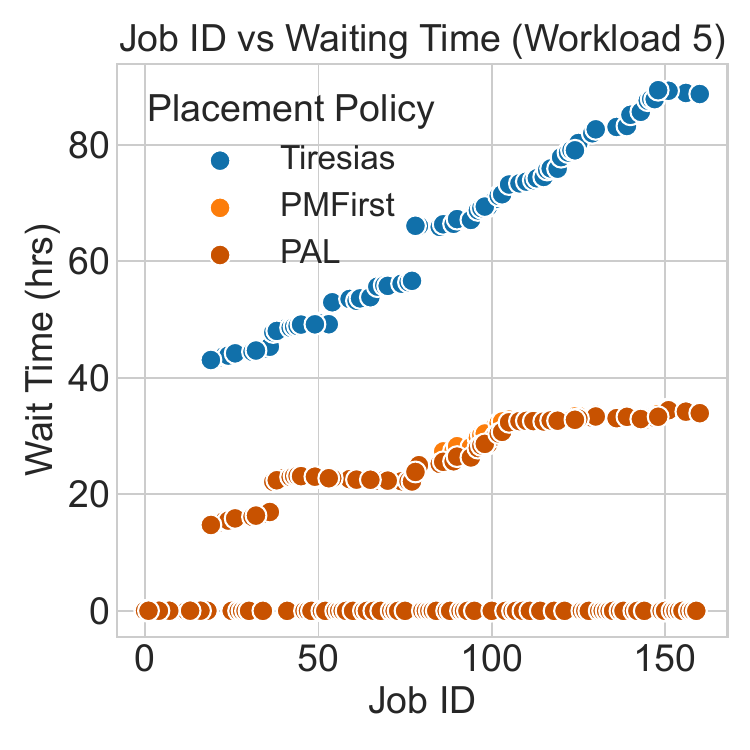}
    \label{subfig:w5}
    \end{subfigure}
    \vspace{-3.5ex}
    \caption{Comparing wait times under different placement policies for workload traces 3 and 5. For brevity we only compare PM-First, PAL, the best-performing baseline (\textbf{Tiresias}).}
    \label{fig:sia-w3w5comparison}
\end{figure}

\noindent
\textbf{PM-First \& PAL}:
Overall, both \textbf{PM-First} and \textbf{PAL} both outperform the baseline placement policies: PM-First improves average JCT by 40\% geomean (min 5\%, max 59\%) while PAL improves average JCT by 43\% geomean (mix 21\%, max 59\%) compared to \textbf{Tiresias}.
\textbf{PM-First} and \textbf{PAL} also improve makespan by 44\% and 47\% respectively over \textbf{Tiresias}.
To better understand \textbf{PM-First}'s and \textbf{PAL}'s range of benefits, we examined the GPU demand distribution for workload traces 3 and 5, which provide the best and the worst improvements, respectively, over \textbf{Tiresias}.
Both workloads have nearly $40$\% single-GPU jobs, but workload 5 also has some very large multi-GPU jobs -- e.g., up to 48 GPU jobs that occupy 75\% of the cluster when they are scheduled and increase the waiting times for subsequent, queued jobs (unsurprisingly, workload 5 has longer wait times in Figure~\ref{fig:sia-w3w5comparison}).
Consequently, workload 5's wait times generally increase for subsequent jobs, as expected with a FIFO scheduler.
For example, in workload 5 an ImageNet job that requests 48 GPUs arrives early (job ID 19) and blocks subsequent jobs from getting sufficient cluster resources, significantly increasing waiting times.
Thus, despite high contention, PAL and PM-First policies more efficiently manage resource allocation and drain the job queue faster than \textbf{Tiresias}, reducing waiting time and providing much larger benefits for workload 5.
Conversely, in workload 3 long-running jobs with large GPU demands only arrive later on (e.g., job ID 60 in a 160 job trace).
As a result, workload 3's jobs have lower wait times, reducing PAL's benefits.

\begin{figure}[bt!]
    \centering
    \includegraphics[width=\linewidth]{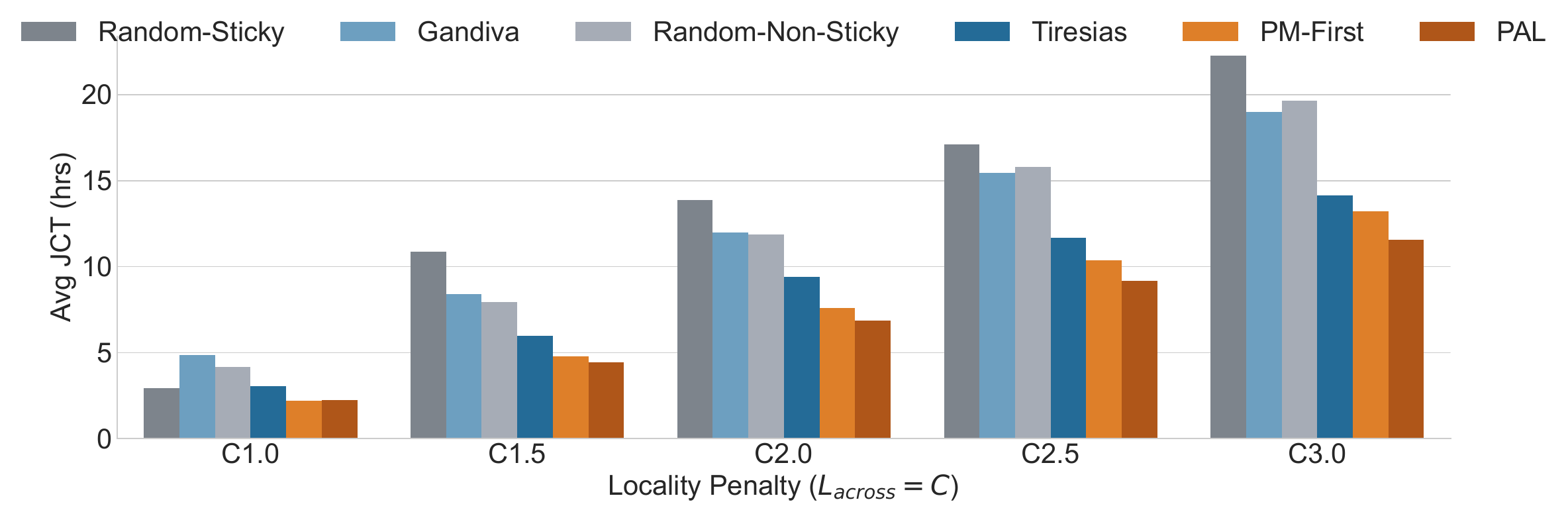}
    \caption{Average JCT for Sia workloads when inter-node locality penalty varies from 1.0 to 3.0.}
    \label{fig:sia-losweep}
\end{figure}

\subsubsection{Varying Locality Penalty}
\label{subsec:sia-philly-eval-localPenal}

Figure~\ref{fig:sia-losweep} shows how various inter-node locality penalty values affect our policies for the Sia-Philly workloads.
As the locality penalty increases, the cost of allocating GPUs across nodes starts to dominate and the placement policies that prioritize packing (\textbf{Tiresias} and \textbf{Gandiva}) start winning over the \textbf{Random} variants.

As the locality penalty increases, the best-performing baseline (\textbf{Tiresias}) improves its average JCT and approaches \textbf{PM-First}'s and \textbf{PAL}'s performance.
For example, %
\textbf{PM-First}'s average JCT improvement over \textbf{Tiresias} decreases from 30\% to 9\% as the locality penalty increases from 1.0 to 3.0.
Unsurprisingly, this demonstrates that \textbf{PM-First} does well when locality penalty is low and \textbf{Tiresias} does well when locality penalty increases.
Nevertheless, even with a large locality penalty, \textbf{PM-First} still outperforms \textbf{Tiresias}, showing the value in harnessing performance variability.
However, by prioritizing both locality and performance variability for multi-GPU jobs (Section~\ref{sec:design}), \textbf{PAL} outperforms both \textbf{PM-First} and \textbf{Tiresias}: as locality penalty is increased from 1.0 to 3.0, \textbf{PAL}'s benefits over \textbf{Tiresias} only decrease from 30\% to 20\% geomean.

\subsection{Synergy Trace Simulations}
\label{subsec:synergy-sim-eval}

\begin{figure}[tb!]
    \centering
    \includegraphics[width=0.8\linewidth]{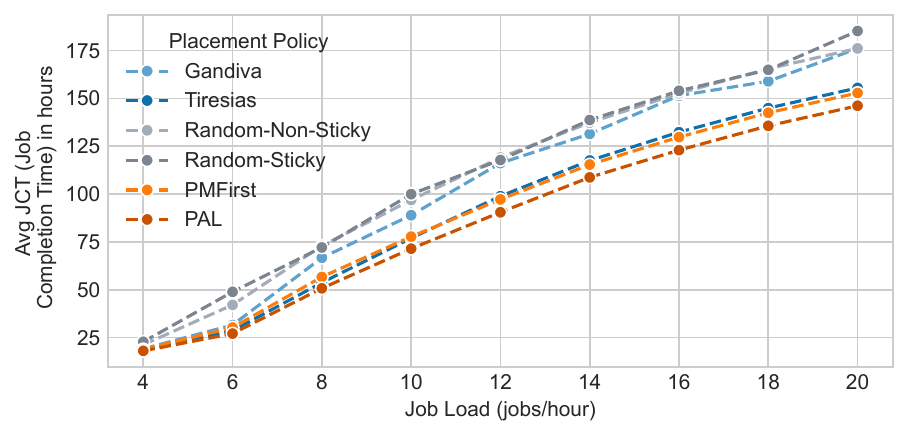}
    \caption{Average JCT for Synergy traces with FIFO scheduler and varying job load.}
    \label{fig:synergy-fifo-jobload}
    \vspace{-1ex}
\end{figure}

For the Synergy traces we vary the job load or arrival rate (jobs/hour) and measure JCTs to evaluate the performance of our policies under varying levels of cluster contention.
Figure~\ref{fig:synergy-fifo-jobload} compares the average JCT at different job loads with various placement policies, for the Synergy traces, on a 256-GPU simulated cluster with a constant locality penalty of $1.7$, and variability profiles drawn from TACC's Longhorn cluster (Figure~\ref{fig:perf_long}).
Unsurprisingly, \textbf{Tiresias} placement again outperforms the other baselines because packing helps avoid the high inter-node locality cost for multi-GPU jobs.
Comparatively, \textbf{PM-First} sees lower improvements with Synergy than Sia: since Synergy has more single GPU jobs (Section~\ref{subsubsec:methodology-appsConfig-sim}), these jobs do not need to be packed.
However, unlike \textbf{PM-First}, \textbf{PAL} successfully prioritizes packed allocations for multi-GPU jobs, where spilling across nodes is prohibitively expensive.
Moreover, \textbf{PAL} \textbf{also} makes the same allocations as \textbf{PM-First} for single GPU jobs that need not be packed.
Thus, \textbf{PAL} still provides the best overall JCT.

Overall, with FIFO scheduling, \textbf{PAL} improves the average JCT from $4\%$ to $9\%$ over \textbf{Tiresias} as job load varies from 4 jobs/hr to 12 jobs/hr, before benefits stabilize around 8\%. 
These benefits happen despite Synergy having $\approx$80\% short-running, single-GPU jobs. 
For Synergy's multi-GPU allocations, the effect of variability is more pronounced -- since these multi-GPU jobs must periodically synchronize across GPUs, their execution is bound by the slowest GPU in their allocation.
Thus, \textbf{PAL} improves the average JCT of multi-GPU jobs by 5\% to 31\% over \textbf{Tiresias} as job load varies from 4 to 12 jobs/hour.
At job loads $> 12$ jobs/hour, multi-GPU jobs see 22\% average JCT benefits with \textbf{PAL}.

\begin{figure}[tb!]
  \centering
  \begin{subfigure}{0.2\textwidth}
    \includegraphics[width=\textwidth]{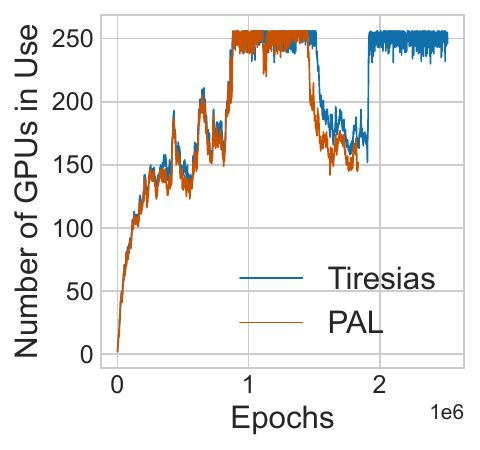}
    \caption{8 jobs/hour}
    \label{fig:first}
  \end{subfigure}
  \begin{subfigure}{0.2\textwidth}
    \includegraphics[width=\textwidth]{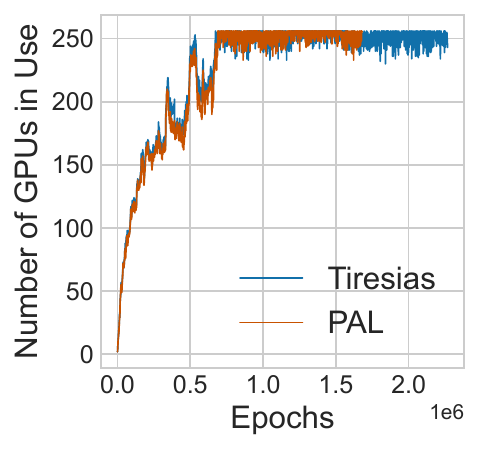}
    \caption{10 jobs/hour}
    \label{fig:second}
  \end{subfigure}
  \caption{GPUs in use for different job loads with \textbf{Tiresias} and \textbf{PAL} placement policies.}
  \label{fig:gpuutil}
  \vspace{-1ex}
\end{figure}

Figure~\ref{fig:gpuutil} shows the number of GPUs in use at every scheduling epoch.
For job loads under 8 jobs/hour, there is low contention and the cluster remains under-utilized, as seen in the utilization dip at around $1.5 \times 10^6$s. 
For 10 jobs/hour, the cluster gets saturated early (at around $0.7 \times 10^6$s) and all 256 GPUs are consistently busy thereafter. 
We observe a similar wait time cascading effect as with the Sia traces, where small execution time improvements enable \textbf{PM-First} and \textbf{PAL} to drain the job queue faster, significantly reducing wait times for subsequent jobs. 
This can be seen in \textbf{PAL}'s utilization pattern, where \textbf{PAL}'s utilization ``runs ahead" of \textbf{Tiresias} (Figure~\ref{fig:gpuutil}), freeing up resources earlier. 
At low contention levels, queuing time for jobs is low. 
\textbf{PAL} and \textbf{PM-First} provide execution time benefits here, but the wait time benefits are limited since most jobs get resources immediately on arrival regardless of placement policy, except for jobs between time $1 \times 10^{6}$ to $1.5 \times 10^{6}$. 

\begin{figure}[tb!]
    \centering
    \includegraphics[width=0.9\linewidth]{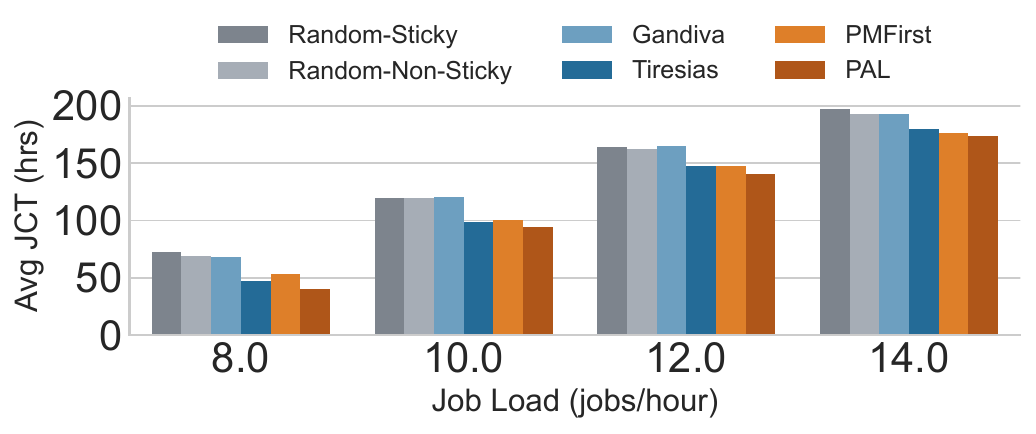}
    \caption{Average JCT for Synergy traces with LAS scheduler and varying job load.}
    \label{fig:synergy-las-jobload}
    \vspace{-1ex}
\end{figure}

\begin{figure}[tb!]
    \centering
    \includegraphics[width=0.9\linewidth]{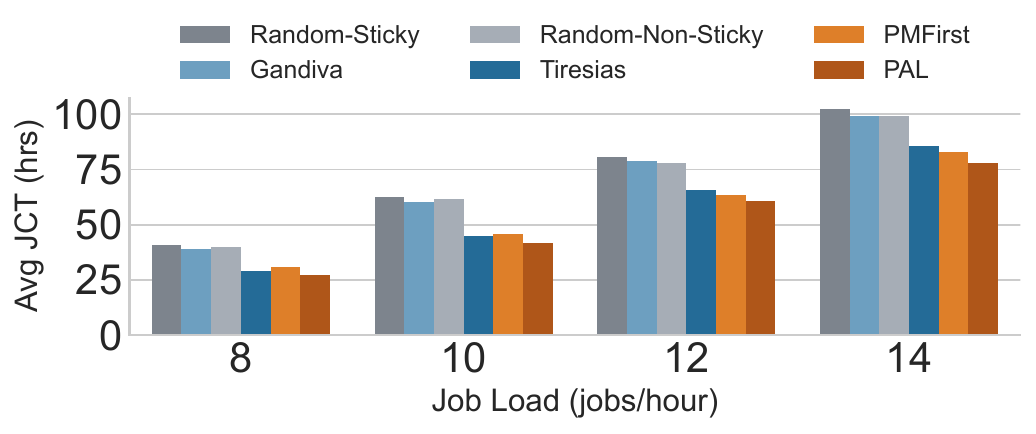}
    \caption{Average JCT for Synergy traces with SRTF scheduler and varying job load.}
    \label{fig:synergy-srtf-jobload}
    \vspace{-1ex}
\end{figure}

\begin{figure}[tb!]
    \centering
    \includegraphics[width=\linewidth]{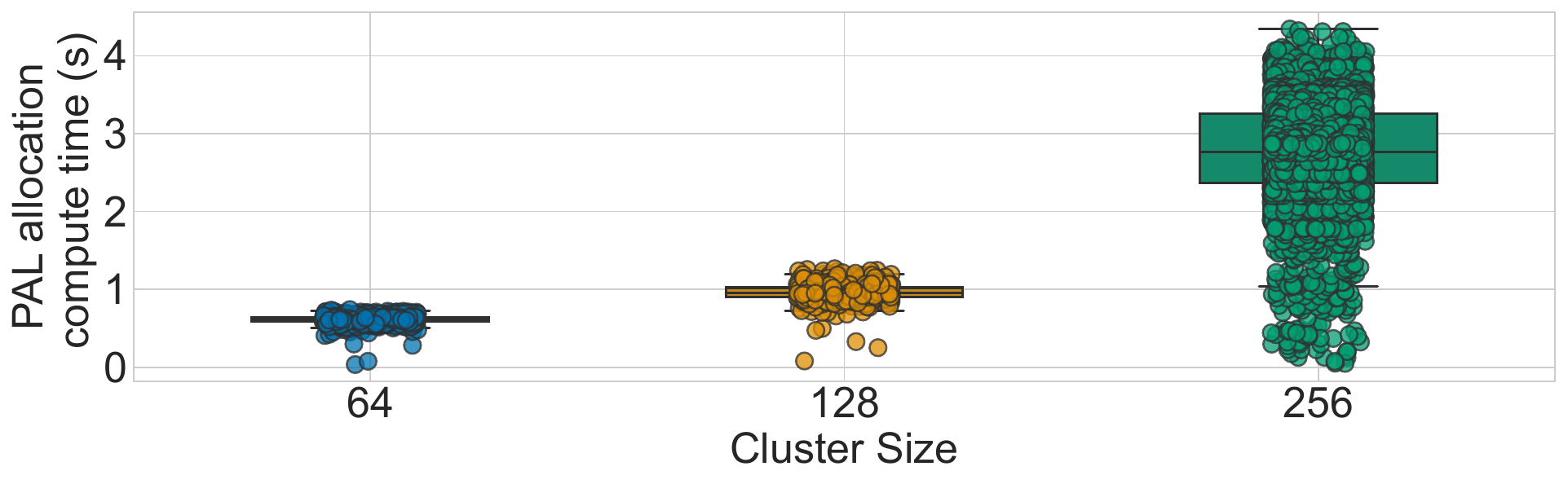}
    \vspace{-1ex}
    \caption{Placement overheads for PAL policy with varying cluster sizes.}
    \label{fig:pal-overheads}
\end{figure}

Finally, to estimate \textbf{PAL} and \textbf{PM-First}'s additional overheads, we measured the time each placement policy takes at every scheduling epoch for different cluster sizes.
Figure~\ref{fig:pal-overheads} shows the distribution of policy computation time over all scheduling epochs, for varying cluster sizes. For a 256-GPU cluster, \textbf{PAL}’s worst-case scheduling time is 4 seconds (for the very first scheduling epoch) with a median of 2.8 seconds.
\textbf{PM-First}'s worst-case computation time is 2 seconds, also for the first epoch.
Thus, both \textbf{PAL} and \textbf{PM-First} complete an epoch’s GPU assignments within 4 seconds -- much smaller than the 300 second epoch duration. 

\subsubsection{Varying Scheduling Policies: SRTF \& LAS}
\label{subsubsec:synergy-sim-eval-varySched}

Figures~\ref{fig:synergy-las-jobload} and~\ref{fig:synergy-srtf-jobload} show variation of average JCT as job loads increase from 8 to 14 jobs/hour for LAS and SRTF schedulers respectively.
With LAS, the absolute values of JCTs are higher because of higher wait times compared to FIFO.
Nevertheless, \textbf{PAL} provides up to 15\% improvement over \textbf{Tiresias}.
With SRTF, \textbf{PAL} provides up to 10\% improvement in average JCT over \textbf{Tiresias}.
These improvements are higher than FIFO's, and can be attributed to differences in wait time patterns between these scheduling policies. 

\begin{figure}[bt!]
  \centering
    \begin{subfigure}{0.15\textwidth}
    \includegraphics[width=\textwidth]{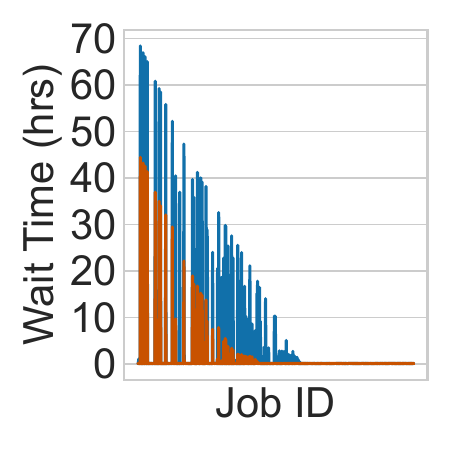}
    \caption{}
  \end{subfigure}
    \begin{subfigure}{0.15\textwidth}
    \includegraphics[width=\textwidth]{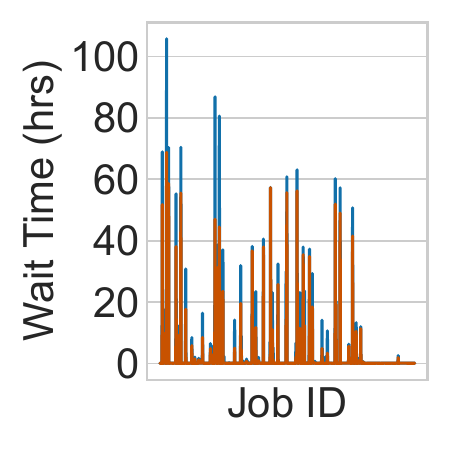}
        \caption{}
  \end{subfigure}
  \begin{subfigure}{0.15\textwidth}
    \includegraphics[width=\textwidth]{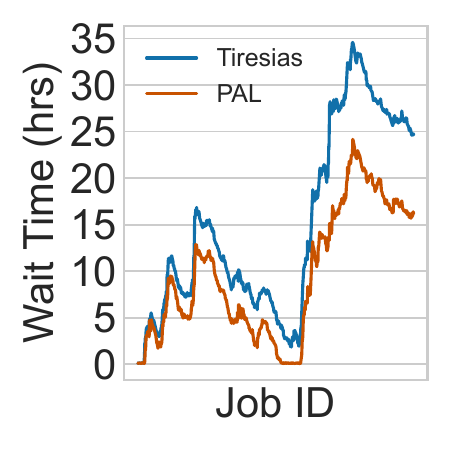}
        \caption{}
  \end{subfigure}
  \caption{Comparing \textbf{Tiresias}'s and \textbf{PAL}'s wait times for Synergy with (a) LAS, (b) SRTF,  and (c) FIFO schedulers.}
  \label{fig:wait-time-comparison}
  \vspace{-2ex}
\end{figure}

Figure~\ref{fig:wait-time-comparison} compares wait times with epochs for the three different schedulers.
For LAS, incoming jobs get higher priority than running jobs since they have no attained service -- decreasing wait time for successive jobs.
The extent of decrease depends on cluster contention levels. 
Figure~\ref{fig:wait-time-comparison}(a) shows this decrease at 8.0 jobs/hr -- wait times go down to 0 for jobs later in the trace. 
Moreover, the absolute values of wait times are large enough to dominate the overall JCT for these jobs. 
\textbf{PAL} often reduces wait time for these long queued jobs, due to its aforementioned run-ahead effect (Section~\ref{subsec:synergy-sim-eval}). 
Conversely, SRTF has fewer wait time spikes than LAS (Figure~\ref{fig:wait-time-comparison}(b)) but still gets some wait time benefits with \textbf{PAL} over \textbf{Tiresias}.
With FIFO scheduling, wait times progressively increase over time. 
Thus, its magnitude of wait times is comparatively lower than either SRTF or LAS.
As a result, \textbf{PAL} correspondingly achieves lower wait time benefits over \textbf{Tiresias} with FIFO scheduling. 

\subsubsection{Varying Locality Penalty}
\label{subsec:synergy-losweep}

\begin{figure}[bt!]
    \centering
    \includegraphics[width=\linewidth]{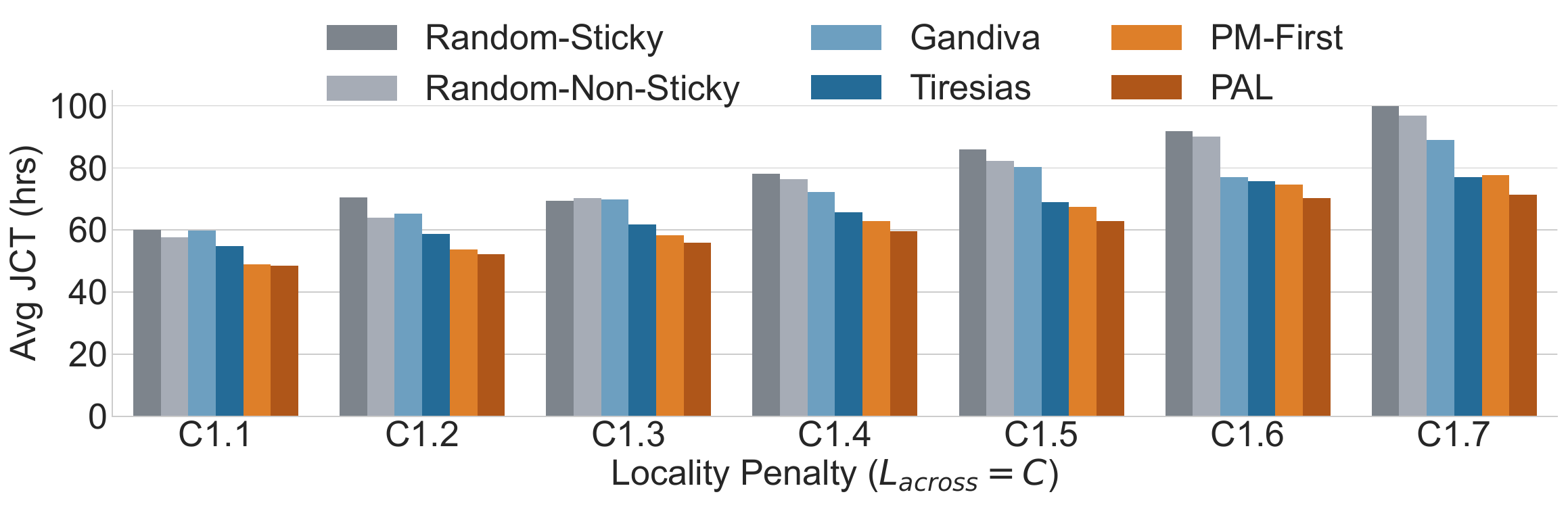}
    \caption{Average JCT for Synergy workload trace when inter-node locality penalty varies from 1.0 to 1.7.}
    \label{fig:synergy-losweep}
\end{figure}

Similar to the Sia workloads in Section~\ref{subsec:sia-philly-eval-localPenal}, Figure~\ref{fig:synergy-losweep} shows how increasing inter-node locality penalty affects our policies for Synergy workloads.
We evaluate the behavior of our policies with the Synergy trace at an arrival rate of 10 jobs/hr, as the locality penalty increases from 1.0 to 1.7.
As with the Sia traces, as locality penalty increases, placement policies that prioritize packing, namely \textbf{Tiresias} and \textbf{Gandiva} achieve lower average JCTs, and the best-performing baseline (\textbf{Tiresias}) approaches \textbf{PM-First}'s and \textbf{PAL}'s performance. 
At a locality penalty of 1.7, both Tiresias and PM-First have nearly the same average JCT. 
However, by prioritizing both locality and performance variability for multi-GPU jobs, \textbf{PAL} continues to outperform both \textbf{PM-First} and \textbf{Tiresias} at higher locality penalty values: as locality penalty increases from 1.0 to 1.7, \textbf{PAL}'s benefits over \textbf{Tiresias} only decrease from 12\% to 7\%. 

%% file: tables/cluster-vs-sim-results.tex
\begin{table}[tb!]
\centering
\caption{Physical cluster \& simulation results.}
\label{tab:cluster-res}
\begin{tabular}{@{}llll@{}}
\toprule
\multirow{2}{*}{\textbf{Placement   Policy}} & \multicolumn{2}{l}{\textbf{Avg JCT (hours)}} & \multirow{2}{*}{\textbf{Cluster-to-Sim Diff}} \\ \cmidrule(lr){2-3}
                                    & \textbf{Cluster} & \textbf{Simulation}                &                                      \\ \midrule
\multicolumn{1}{l|}{Tiresias}  & 1.76    & \multicolumn{1}{l|}{1.56} & 11\%                               \\
\multicolumn{1}{l|}{PAL}            & 1.35     & \multicolumn{1}{l|}{1.16} & 14\%                               \\ \midrule
\% Improvement                      & 24\%  & 26\%                      &                                      \\ \bottomrule
\end{tabular}
\vspace{-1ex}
\end{table}

%% file: relWork.tex
\section{Related Work}
\label{sec:relWork}

\noindent
\textbf{CPU Variability}: CPU-based systems have proposed and adopted several solutions to handle variability, but largely focus on different aspects or scales than ours, making them complementary to our proposed approaches.
Dynamic load balancing algorithms~\cite{MenonAcun2013-thermAwareLoadBal,AcunKale2016-dynLoadBal,Acun-hipc17-proactiveCooling} use adaptive runtime systems that estimate a task’s completion time if it is moved from one processor to another for single CPU experiments. Speed-aware solutions identify overloaded nodes as ones with throttled core frequencies~\cite{AcunKale2016-dynLoadBal} while thermal-aware solutions identify hotspots~\cite{MenonAcun2013-thermAwareLoadBal,Acun-hipc17-proactiveCooling}. Conversely, we profile and directly use performance variability for single- and multi-GPU jobs, which requires managing placement and locality across multiple devices for a given job. Other solutions~\cite{Chasapis-RuntimeGuided-ICS16} minimize variability within a NUMA node using a runtime-guided search for thread and task socket assignments. However, this does not scale to GPUs with thousands of threads or larger systems with hundreds/thousands of GPUs. Some solutions estimate total consumed power when scheduling jobs to stay within a system-wide power budget. To estimate power variability, jobs either use static profiling tables~\cite{Inadomi-SC15} or power prediction models~\cite{Chasapis-RuntimeGuided-ICS16}. While these solutions are power variability aware, neither consider it in their placement policies.

\noindent
\textbf{GPU Performance Variability}:
Prior VLSI and architecture works have developed hardware-level techniques and optimizations to mitigate or tolerate process variation either within a single GPU or components of a single GPU such as compute units, memories, or the register file~\cite{Tan-Gpuvar-2019,Tan-RegFileVar-2016,Maura-mupmusys-2023}.
However, these solutions are local to the GPU and do not extend to multi-GPU systems.
Moreover, they also focus on process variation and do not address dynamic variability due to PM and the overall effect of variability on application performance like our work.

\noindent
\textbf{Cluster Schedulers}:
Unlike batch scheduling mechanisms such as SLURM~\cite{yoo-slurm-2003}, ML cluster schedulers for GPU-rich systems are tailored towards the unique demands and constraints imposed by ML training, such as checkpointing and periodic job migration.
In this work, we specifically focus on such round-based, preemptive schedulers for ML training in large-scale GPU clusters.
These schedulers make GPU allocation decisions for competing ML jobs with specific scheduling objectives.
Scheduling is an extremely active area of research, with over 23 schedulers proposed for ML training on GPU datacenters since 2022.
Thus, there are numerous prior works, as comprehensively summarized by Gao et al.~\cite{gao2022-survey}.
However, the most relevant ones to our work have developed schedulers to optimize for different objectives, such as JCT, utilization, cost, fairness, and deadlines~\cite{Subramanya-Sia-Sosp23,Qiao-Pollux-OSDI21,Narayanan-pop-sosp21,Gu-Tiresias-NSDI19,Hu-GangScheduling-SC21,Mahajan-Themis-OSDI20, Narayanan-Gavel-osdi20,Bian-EvolveBsz-SC21,Hu-Helios-SC21}.
Sia~\cite{Subramanya-Sia-Sosp23}, ONES~\cite{Bian-EvolveBsz-SC21} and Helios~\cite{Hu-Helios-SC21} perform scheduling at the same cluster scale as PAL.
Both Sia and ONES proposed algorithms for adaptive batch size scaling of jobs, while Helios derived insights from job submission data to come up with a Quasi Shortest Service First scheduler.
However, unlike PAL, all these schedulers are either performance variability agnostic or, like Gavel~\cite{Narayanan-Gavel-osdi20}, only consider heterogeneity across different GPUs from the same vendor in an HPC system.
In comparison, PM-First enables fine-grained iso-architecture variability that allows them to make variability-informed decisions. 
Moreover, PAL co-optimizes variability with packing to further improve performance. 

\noindent
\textbf{Placement Policies}:
Jeon et al.~\cite{Jeon-Philly-2019} %
recommended that multi-tenant GPU cluster schedulers should prioritize locality to reduce distributed communication costs for long-running jobs. %
Accordingly, several schedulers try to pack jobs onto fewer nodes~\cite{Gu-Tiresias-NSDI19, Mahajan-Themis-OSDI20, Narayanan-Gavel-osdi20, Subramanya-Sia-Sosp23}. 
For example, Tiresias profiles a job to identify its sensitivity to placement and performs packed placement for jobs with high communication overheads~\cite{Gu-Tiresias-NSDI19}.
However, all these schedulers are agnostic to GPU variability, which we show in Section~\ref{sec:eval} causes PAL to outperform them. 
Gavel~\cite{Narayanan-Gavel-osdi20} and Sia~\cite{Subramanya-Sia-Sosp23} consider different accelerator architectures on heterogeneous clusters, while Amaral et al.~\cite{Amaral-Topology-SC17} consider heterogeneity in GPU interconnect topology.
These solutions also assume that all GPUs of a given architecture deliver equal performance.
Thus, to the best of our knowledge, our work is the first to create cluster schedulers aware of performance variability among GPUs in HPC systems with the same architecture and make scheduling decisions informed by this variability.

%% file: 08_conclusion.tex
\section{Conclusion}
\label{sec:conclusion}

GPU-rich clusters are becoming increasingly prevalent to meet the computing needs of large-scale ML workloads.
However, performance variability can significantly affect their cluster performance, utilization, and load balance -- trends that are likely to worsen as ML algorithms continue to scale.
Thus, ML schedulers in large scale systems must embrace and harness this variability. 
Accordingly, we propose \textbf{PAL}, %
which 
uses application-specific variability characterization to intelligently perform variability-aware GPU allocation.
Moreover, PAL %
co-optimizes job placement for both variability and locality to reduce communication overheads.
Overall, across a mix of ML workloads, our evaluation shows that PAL improves on state-of-the-art ML cluster schedulers across a number of metrics.
Moreover, we expect HPC and HPC+ML workloads will exhibit similar benefits.

%% file: acks.tex
\section*{Acknowledgment}

We thank the anonymous shepherd and the SC reviewers for their constructive comments and suggestions that improved this work.
We express our gratitude to Saurabh Agarwal and Song Bian (University of Wisconsin-Madison) for their support in enabling Blox on Texas Advanced Computing Center's (TACC) Frontera cluster.
We also thank Zhao Zhang for providing us machine access on the TACC clusters we used and helping us run some of the experiments. 
This work is supported by NSF grant CNS-2312688, NSF grant CCRI-AAG9851, and a TACC allocation.
Sinclair has an affiliate appointment with AMD Research.